\documentclass[apj, 12pt, letterpaper]{emulateapj}
\usepackage{apjfonts}

\begin{document}

\slugcomment{\bf}
\slugcomment{Accepted for the Astrophysical Journal}

\title{Dynamics and Disequilibrium Carbon Chemistry in Hot Jupiter
Atmospheres, With Application to HD 209458b} 
\shorttitle{Carbon Chemistry on HD 209458b}
\shortauthors{Cooper \& Showman}

\author{Curtis S.\ Cooper\altaffilmark{1} and Adam P. Showman\altaffilmark{1}}

\altaffiltext{1}{Department of Planetary Sciences and Lunar and Planetary
Laboratory, The University of Arizona, 1629 University Blvd., Tucson, AZ 85721 USA; 
curtis@lpl.arizona.edu, showman@lpl.arizona.edu}

\begin{abstract}
\label{Abstract}

Chemical equilibrium considerations suggest that, assuming solar elemental
abundances, carbon on HD 209458b is sequestered primarily as carbon monoxide
(CO) and methane ($\rm CH_4$).  The relative mole fractions of CO(g) and $\rm
CH_4$(g) in chemical equilibrium are expected to vary greatly according to
variations in local temperature and pressure.  We show, however, that in the
$p$ = 1--1000 mbar range, chemical equilibrium does not hold.  To explore
disequilibrium effects, we couple the chemical kinetics of CO and $\rm CH_4$
to a three-dimensional numerical model of HD 209458b's atmospheric
circulation.  These simulations show that vigorous dynamics caused by uneven
heating of this tidally locked planet homogenize the CO and $\rm CH_4$
concentrations at $p < 1$ bar, even in the presence of lateral temperature
variations of $\sim\!500$--1000 K.  In the 1--1000 mbar pressure range, we
find that over 98\% of the carbon is in CO.  This is true even in cool regions
where $\rm CH_4$ is much more stable thermodynamically.  Our work shows
furthermore that planets 300--500 K cooler than HD 209458b can also have
abundant CO in their upper layers due to disequilibrium effects.  We
demonstrate several interesting observational consequences of these results.

\end{abstract}



\section{Introduction}
\label{Introduction}

\citet{Charbonneau:2000} and \citet{Henry:2000} discovered the first known
transiting extrasolar giant planet (EGP), HD 209458b.  Transits are possible
if the orbital plane of the planet is tilted along our direct line of sight.
The detection of HD 209458b in transit has been a watershed in the study of
planets around other stars because physical properties of the planet can be
inferred directly from the transit data.  The planet's orbit is nearly
circular, with a period of 3.5257 days \citep{Laughlin:2005b}.  Furthermore,
\citet{Laughlin:2005a} have combined transit data with high-resolution spectra
of the star to infer accurate values for the planet's mass and radius: $M_{\rm
p} = (0.69 \pm 0.05)\,M_{\rm Jupiter}$ and $R_{\rm p} = (1.32 \pm
0.05)\,R_{\rm Jupiter}$.  

Photons from HD209458b and the transiting planet TrES-1 have recently been
detected \citep{Deming:2005a, Charbonneau:2005}.  In addition, transmission
spectroscopy measurements, in which observers track changes in the star's
spectral features while the planet is in transit, have unveiled key features of
the composition of HD 209458b's atmosphere \citep{Brown:2001a, Brown:2001b,
Charbonneau:2002, Vidal-Madjar:2003, Vidal-Madjar:2004}.  The more recent
detections of transiting planets OGLE-TR-56b \citep{Konacki:2003}, OGLE-TR-113b
and 132b \citep{Bouchy:2004}, TrES-1 \citep{Alonso:2004}, OGLE-TR-111b
\citep{Pont:2004}, OGLE TR-10b \citep{Bouchy:2005a, Konacki:2005}, HD 149026b
\citep{Sato:2005}, and HD 189733b \citep{Bouchy:2005b} reinforce the importance
of the transit method for the characterization of close-in EGP atmospheres.

HD 209458b resides in close proximity (0.046 AU) to a Sun-like star in the
constellation Pegasus.  The intense stellar heating of close-in EGPs drives
global-scale atmospheric dynamics.  As shown by \citet{Showman:2002},
\citet{Cho:2003}, \citet{Burkert:2005}, and \citet{Cooper:2005}, winds of
$\sim\!3000\;\rm m\;s^{-1}$ carry material across a hemisphere in only
$\tau_{\rm dyn,\,h} \sim 10^5\;\rm sec$.  Vertical transport is rapid as well,
with typical wind speeds $w \sim 20\rm\;m\;s^{-1}$ at $p = 1$ bar (see
\S \ref{section:Analysis_Interpretation}). 

Theory predicts that HD 209458b is locked by strong stellar tides to a
synchronously rotating state, owing to the short time for spin-orbit
synchronization of a planet at 0.046 AU \citep[see e.g.,][]{Barnes:2003}.  The
planet therefore receives stellar radiation on one hemisphere only
\citep{Guillot:1996, Showman:2002}.  Various effects causing deviations from
synchronous rotation have been explored in the literature, such as non-zero
eccentricity or substantial obliquity pumped by the presence of another planet
\citep{Laughlin:2005a, Winn:2005}.  However, the apparent lack of a third
massive body in the system \citep{Laughlin:2005b}, as well as the timing of
the transit and secondary eclipse \citep{Deming:2005a}, makes a large
eccentricity or obliquity unlikely for this system.  

We study the atmosphere under the assumption of synchronous rotation.  We
note, however, that deviations from a completely synchronous state are
conceivable.  For example, effects of atmospheric tides are important on
Venus, as shown by \citet{Correia:2001}.  In an order-of-magnitude
calculation, \citet{Showman:2002} suggested that atmospheric dynamics could
induce deviation from synchronous rotation by up to a factor of 2.  A factor
of 2 is the extreme case; 10--20\% deviation might be more typical.  This
estimate depends on the planet's tidal dissipation factor ($Q$), as well as
properties of the circulation itself and the processes transferring angular
momentum and rotational kinetic energy from the atmosphere to the interior,
which are highly uncertain.

The G0V primary irradiates the planet up to an extremely high effective
temperature \citep[$T_{\rm eff}\sim1340$ K; see][]{Iro:2005} compared to the
planets of our Solar System.  This leads to very different chemistry than we
are familiar with in planetary atmospheres.  The atmosphere is primarily
composed of hydrogen and helium.  Carbon, nitrogen, and oxygen are all minor
constituents.  Despite their low abundances, these elements are important for
determining the spectral characteristics of the planet because they form
molecules having strong spectral signatures \citep{Seager:1998}.
Radiative-equilibrium calculations by \citet{Iro:2005} show that in the 1 mbar
to 10 bar pressure regime, which is the region of primary interest for
observations, the temperature lies in the 1000--2000 K range.  At these
temperatures and pressures, chemical equilibrium considerations suggest that
carbon on HD 209458b is sequestered primarily in carbon monoxide (CO) and
methane ($\rm CH_4$) \citep{Burrows:1999, Lodders:2002}.  These models show
that CO(g) is the most stable carbon-bearing species at high temperatures and
low pressures.  At low temperatures and high pressures, $\rm CH_4$(g) is more
stable thermodynamically.  The oxygen not bound up in rock is present as
either CO(g) or water vapor, $\rm H_2O$(g).

In planetary atmospheres, however, chemical equilibrium may not hold.  There
are several reasons for this.  First, photochemical dissociation can elevate
the concentrations of photochemical products above equilibrium levels
\citep{Moses:2000}.  \citet{Prinn:1977} note that photochemical oxidation of
$\rm CH_4$ is a well-known source of CO in the Earth's atmosphere.  Second,
impacts by meteorites, comets, etc., deposit metals and other compounds into
the atmospheres of giant planets \citep[see, e.g.,][]{Prather:1978,
Moses:1992}.  Third, chemical reactions may not always occur quickly enough
relative to dynamical mixing for chemical equilibrium to be attained
\citep{Prinn:1977, Smith:1998}.

\citet{Liang:2004} show that the first effect---photochemical production---is
unlikely to significantly alter the concentration of CO in HD 209458b's
stratosphere ($p < 1$ mbar).  It is unknown how much impactors have modified
the composition of HD 209458b's atmosphere during its evolution.  We note,
however, that calculations by \citet{Bezard:2002} show that CO deposition by
impactors cannot explain the observed abundance of CO ($\sim\!1$ ppb, as
opposed to essentially zero in equilibrium) in Jupiter's lower troposphere.
On Jupiter, rather, disequilibrium results from convective upwelling from deep
regions ($p\sim\!300$ bar), where the equilibrium abundance of CO is higher.
As the reaction rates plummet with decreasing temperature and pressure in
ascending fluid parcels, the CO concentration quenches to its observed value.  

Photospheric pressures lie in the 10--500 mbar range for all IR wavelengths
\citep{Fortney:2005a, Burrows:2005, Barman:2005, Seager:2005}.  If large
temperature variations of 500--1000 K in this region are a reality on HD
209458b \citep{Showman:2002, Cho:2003, Burkert:2005, Cooper:2005}, it is
conceivable that chemical quenching occurs in the horizontal rather than the
vertical direction.  In this hypothetical scenario, fluid parcels attain
chemical equilibrium on the planet's dayside, where the temperature is high
and the reaction kinetics are fast.  Vigorous dynamical motions then carry
these equilibrated parcels to the nightside, where they rapidly cool to
temperatures where the reactions are too slow to equilibrate the system.
Depending on the meteorology and variations of the chemical rates with
temperature and pressure, complex lateral CO distributions could potentially
result.  For example, it would then be possible for CO to be abundant at the
evening limb but depleted at the morning limb.

To test the hypothesis that disequilibrium processes modify the abundances of
CO and $\rm CH_4$ in the observable atmosphere of HD209458b, we couple the
reaction kinetics to a three-dimensional nonlinear model of the atmospheric
dynamics.  We show that at the temperatures and pressures relevant for
observations of HD 209458b's atmosphere, the timescale for CO/$\rm CH_4$
interconversion is everywhere long:~$\tau_{\rm chem} \gg 10^5$ s.
Disequilibrium results in regions where the timescale for advection of CO by
dynamics, $\tau_{\rm dyn}$, is shorter than $\tau_{\rm chem}$.  Due to
uniformly large values of $\tau_{\rm chem}$ for $p < 1$ bar, the simulations
definitively show that quenching occurs in the vertical, not the horizontal.
As on Jupiter, we find that the concentration of CO for $p < 1$ bar depends on
its value deeper in the atmosphere.  Vertical quenching and efficient
horizontal mixing lead to a high-abundance, homogeneous distribution of CO at
the photosphere.  As a result, in cool regions (e.g., the nightside), the CO
concentration exceeds equilibrium values by many orders of magnitude.  

Curiously, CO was not detected on HD 209458b in a recent ground-based search
using transmission spectroscopy at $2\,\micron$ \citep{Deming:2005b}.  Several
possible reasons for the CO null detection have been noted by the authors and
others.  First, CO may not be abundant in this planet's atmosphere, though
this is only possible if the planet is significantly cooler (by $\sim\!500$ K)
than predicted by current radiative-equilibrium models \citep{Burrows:2004,
Chabrier:2004, Iro:2005}.  Second, CO may present in the visible atmosphere in
some regions but depleted at the limb \citep{Iro:2005}.  Third, the spectral
signature of CO at the planet's limb may be obscured either due to
photochemical haze or a high-altitude silicate cloud \citep{Fortney:2003,
Burrows:2004}.  This explanation is favored by \citet{Deming:2005b}.  It
should be noted, however, that the CO null detection reported by
\citet{Deming:2005b} relies on the radiative-equilibrium calculations of
\citet{Sudarsky:2003}.  If unexpected opacity sources (though not necessarily
clouds) play a strong role in the formation of spectral lines near $\rm
2\;\micron$, an upper limit to the CO abundance cannot be made with high
confidence based on these observations.  We note $\rm CH_4$ has not been seen
on this planet either, despite attempts to detect its signature
\citep{Richardson:2003a, Richardson:2003b}.

We review in \S \ref{section:Equilibrium_Considerations} the theoretical
reasons to expect CO in high abundance in HD 209458b's atmosphere based on
chemical equilibrium arguments.  We discuss in
\S \ref{section:Chemical_Timescales} the main chemical processes involved and
derive the timescale for CO/$\rm CH_4$ interconversion.  In
\S \ref{section:Model}, we describe our coupling of chemistry to the
atmospheric dynamics model of \citet{Cooper:2005}.  We present the results of
these simulations in \S \ref{section:Simulation_Results}.  We explain and
interpret these results in \S \ref{section:Analysis_Interpretation}, comparing
our simulations to a simpler view of quenching presented by
\citet{Smith:1998}.  We enumerate the major uncertainties in these
calculations in \S \ref{section:Discussion}.  In \S \ref{section:Implications},
we discuss implications of this work for near-future observations of HD
209458b and other close-in EGPs.

\section{Chemical Equilibrium Considerations}
\label{section:Equilibrium_Considerations}

Prior discussions of chemistry on EGPs are based on chemical equilibrium
models \citep{Burrows:1999, Lodders:2002}.  Solar abundances of the elements
\citep{Anders:1989, Lodders:1998} are generally assumed in these schemes,
which perform global minimization of the Gibbs free energy of the system,
often including thermochemical data for hundreds of compounds.  This
minimization algorithm allows the concentrations of all chemicals to be solved
for simultaneously.  In such thermochemical calculations, once the elemental
abundances in the atmosphere have been specified, the abundances of all
chemical species in the system depend solely on the local temperature and
pressure (allowing for simple ideas like rainout).

Simulations by \citet{Cooper:2005} suggest that the temperature at the
photosphere of HD 209458b varies across the globe from $\sim\!1000$--1500 K.
Chemical equilibrium models show that at these temperatures, the atmosphere of
this planet should contain appreciable quantities of CO and $\rm CH_4$.
\citet{Lodders:2002} show that HD 209458b's atmosphere straddles the line of
equal CO/$\rm CH_4$ abundance, in the absence of disequilibrium effects.  The
relative concentrations of CO and $\rm CH_4$ are strong functions of local
temperature and pressure.  At 1 bar, where $\rm T \la 1150$ K, $\rm CH_4$
is the dominant carbon-bearing species (in chemical equilibrium), while CO is
more abundant at this pressure for higher temperatures (T > 1150 K). 

In the temperature and pressure ranges of interest for this investigation,
calculation of the equilibrium abundances of CO and $\rm CH_4$ is
straightforward and reduces to several simple algebraic expressions.  These
expressions are used in our disequilibrium model, in which the mole fraction
of CO is relaxed to its value in chemical equilibrium (see
\S \ref{subsection:Model_CO_Chemistry}).  The expressions, as we will show,
include the metallicity dependence directly.  This allows us to explore the
sensitivity of our disequilibrium calculations to the metallicity of HD
209458b's atmosphere, which is conceivably enhanced from that of the host
star.  Such enhancement is seen on Jupiter, where C, N, and probably O are
several times more abundant relative to hydrogen than at the photosphere of
the Sun \citep{Atreya:1999}.

We define $X_{\rm i}$ to be the mole fraction of the i-th species in the
system; e.g., $X_{\rm H_2}$ is the mole fraction of molecular hydrogen.  It is
given by the ratio of the partial pressure of the i-th species to the total
atmospheric pressure (i.e., $X_{\rm i} = p_{\rm i}/p$).  At the temperatures
and pressures in this region of HD 209458b's atmosphere, hydrogen is primarily
molecular $\rm H_2$ [J. Fortney, private communication].  We take $X_{\rm H_2}
= 0.83$ for the $\rm H_2$ mole fraction throughout this paper (i.e., we assume
that $\rm H_2$ dissociation is everywhere negligible).

Although oxygen is mostly in CO and $\rm H_2O$, some oxygen is also
sequestered in rock, most of which forms deeper than 1 bar
\citep{Lodders:2002, Fortney:2005a}.  We assume 16\% of the atmospheric oxygen
(the maximum possible in a solar abundance mixture is in silicates.  Depending
on where in the atmosphere $\rm MgSiO_3$ condenses, this may underestimate (by
$\sim\!5$\%) the oxygen available to affect CO/$\rm CH_4$ chemistry in the
region of interest.

Defining $c_1$ and $c_2$ as the mixing ratios of gaseous carbon and oxygen in
the atmosphere, the mole fractions $X_{\rm CO}$, $X_{\rm CH_4}$, and $X_{\rm
H_2O}$ are related through mass balance:
\begin{equation}
\label{equation:c1}
X_{\rm CH_4} + X_{\rm CO} = c_1 = 5.91\times 10^{-4}
\end{equation}
\begin{equation}
\label{equation:c2}
X_{\rm CO} + X_{\rm H_2O} = c_2 = 1.04\times 10^{-3}
\end{equation}
for solar metallicity.  We see from equations \ref{equation:c1} and
\ref{equation:c2} that water vapor will always be present in the system,
even if 100\% of the carbon is in CO, since oxygen is more abundant than
carbon.

The third expression relating the quantities $X_{\rm CO}$, $X_{\rm CH_4}$, and
$X_{\rm H_2O}$ derives from the equilibrium expression of the net
thermochemical reaction 
\begin{equation}
\label{reaction:CO_CH4_H2O_equil}
\rm CO \, + \, 3H_2 \, \leftrightarrow \, CH_4 \, + \, H_2O.
\end{equation}
The equilibrium expression for reaction \ref{reaction:CO_CH4_H2O_equil}, which
we denote as $\rm \mathnormal{K}_{\ref{reaction:CO_CH4_H2O_equil},\,\rm eq}$,
is
\begin{equation}
\label{equation:CO_CH4_H2O_equil}
\frac{X_{\rm CO} X_{\rm H_2}^3 p^2}{X_{\rm CH_4} X_{\rm H_2O}} = \rm
exp\left [-\frac{\Delta_f \mathnormal{G}_{\rm CO} - \Delta_f \mathnormal{G}_{\rm
CH_4} - \Delta_f \mathnormal{G}_{\rm H_2O}}{\mathnormal{RT}} \right ]
= \mathnormal{K}_{\ref{reaction:CO_CH4_H2O_equil},\,\rm eq}.
\end{equation}

In equation \ref{equation:CO_CH4_H2O_equil}, $p$ is atmospheric pressure
[bar], $T$ is the temperature [K], and $\rm \mathnormal{R} =
8.314\;J\;mol^{-1}\;K^{-1}$ is the universal gas constant.  The symbols $\rm
\Delta_f \mathnormal{G}_{\rm CO}$, $\rm \Delta_f \mathnormal{G}_{\rm CH_4}$,
and $\rm \Delta_f \mathnormal{G}_{\rm H_2O}$ represent the \em Gibbs free
energies of formation \em of CO, $\rm CH_4$, and $\rm H_2O$, respectively.
The Gibbs free energies of formation depend on temperature sensitively but
only weakly on pressure.  We simply use the values at standard pressure (1
bar), which are tabulated in the \em NIST-JANAF Thermochemical Tables \em
\citep{Chase:1998}.

With the three equations \ref{equation:c1}, \ref{equation:c2}, and
\ref{equation:CO_CH4_H2O_equil}, we can solve for the three equilibrium mole
fractions $X_{\rm CO}$, $X_{\rm H_2O}$, and $X_{\rm CH_4}$ as a function of
atmospheric pressure ($p$) and temperature ($T$).  Defining $\mathnormal{f} =
K_{\ref{reaction:CO_CH4_H2O_equil},\,\rm eq} / (p^2 X^3_{\rm H_2} )$, we solve
for $X_{\rm CO}$.  We take the negative root of the quadratic equation, which
represents the physical solution:
\begin{equation}
\label{equation:X_CO}
X_{\rm CO} = \frac{-b - \sqrt{b^2 - 4ac}}{2a},
\end{equation}
with $a = f$, $b = -f\,(c_1 + c_2) - 1$, and $c = f\,c_1c_2$.  The other 
concentrations are then easily solved for
using the mass balance expressions (equations \ref{equation:c1} and
\ref{equation:c2}): $X_{\rm CH_4} = c_1 - X_{\rm CO}$ and $X_{\rm H_2O} = c_2
- X_{\rm CO}$.  The solution is valid for temperatures between 500--2500 K
over a broad range of pressures in the atmosphere.  It is also valid for
atmospheres with enhanced or depleted C and O abundances by simply tuning
$c_1$ and $c_2$ (so long as $c_1$ and $c_2$ are kept small: $c_1,\,c_2\ll
0.1$).

We show in Table \ref{table:CO_fraction} the ratio of CO to total carbon in
the atmosphere ($X_{\rm CO}\,/\,c_1$), assuming chemical equilibrium, for a
variety of temperatures and pressures.  The table shows that the CO
abundance---assuming equilibrium conditions---is a strong function of pressure
and especially temperature.  If large temperature variations in the atmosphere
are indeed a reality on HD 209458b, as predicted both in radiative-equilibrium
models \citep[e.g., ][]{Iro:2005} and simulations of the atmospheric dynamics
\citep{Cooper:2005}, large gradients in the equilibrium concentrations of CO
and $\rm CH_4$ result.  We now consider disequilibrium effects, which lead to
vastly different conclusions about the relative concentrations of CO and $\rm
CH_4$ in the upper layers of this planet's atmosphere.

\section{Chemical Pathways for CO Reduction}
\label{section:Chemical_Timescales}

CO concentrations above equilibrium values have been discovered both on
Jupiter \citep{Beer:1975, Larson:1978} and on the brown dwarf Gl 229B
\citep{Noll:1997, Oppenheimer:1998, Saumon:2000}.  \citet{Fegley:1996} and
\citet{Griffith:1999} show that CO disequilibrium in Gl 229B results from the
inefficiency of chemical reduction of CO in regions of low temperature and
pressure relative to the rate of vertical mixing.  The CO seen in the spectrum
of Gl 229B originates from deeper in the atmosphere, where CO is abundant in
chemical equilibrium.  

Similarly, \citet{Prinn:1977} suggest that the CO detected in the troposphere
of Jupiter was transported up by convection from deeper, hotter layers of the
planet.  \citet{Bezard:2002} explore several explanations for the unexpected 1
ppb CO concentration observed on this planet.  They conclude that convective
mixing from deeper layers remains the most likely source of Jupiter's
tropospheric CO, although the reactions suggested by \citet{Prinn:1977} are
now thought to be too slow to explain the observed CO abundance.

We hypothesize that dynamics may drive the abundance of CO on HD 209458b away
from chemical equilibrium values.  To ascertain the extent to which dynamics
perturbs chemical equilibrium in this atmosphere, we first calculate the
timescale for the processes converting CO to $\rm CH_4$ and back.  This
depends on the chemical reactions involved.  Two independent reaction pathways
for the chemical reduction of CO have been considered in the literature.  

The following reaction was suggested by \citet{Prinn:1977}:
\begin{equation}
\label{reaction:Prinn}
\rm H_2\,+\,H_2CO \rightarrow OH\,+\,CH_3
\end{equation}
This reaction is the rate-limiting step in a three-part sequence resulting in
the chemical reduction of CO to $\rm CH_4$.  These authors calculate a rate
constant for reaction \ref{reaction:Prinn} of
\begin{equation}
\label{equation:Prinn_rate}
\rm \mathnormal{k}_{\ref{reaction:Prinn}} = 2.3\times 10^{-10}\,
exp\left(-\frac{36,200\:K}{\mathnormal{T}}\right)\;cm^3\;s^{-1}.
\end{equation}

\citet{Prinn:1977} note that the other steps leading to the reduction of CO to
$\rm CH_4$ are fast relative to reaction \ref{reaction:Prinn}.  They use
equation \ref{equation:Prinn_rate} to estimate the chemical lifetime of CO
in disequilibrium as a function of temperature and pressure.

Based on more recent measurements, however, \citet{Griffith:1999} discuss that
the reaction rate for reaction \ref{reaction:Prinn} is significantly slower
than the rate assumed by \citet{Prinn:1977}:
\begin{equation}
\label{equation:Prinn_modified_rate}
\rm \mathnormal{k}_{\ref{reaction:Prinn},\:correct} = 3.4\times
10^{-8}\,\left(\frac{\mathnormal{T}}{K}\right)^{-1.12}
\,exp\left(-\frac{43,192\:K}{\mathnormal{T}}\right)\;cm^3\;s^{-1}.
\end{equation}

Furthermore, as shown by \citet{Smith:1998}, the convective timescale used by
\citet{Prinn:1977} to derive the quench level is incorrect because it does not
account for the gradient in both the lifetime and equilibrium abundance of CO
over an atmospheric scale height.  Using the updated reaction rate, equation
\ref{equation:Prinn_modified_rate}, and the correct timescale for convective
mixing, \citet{Bezard:2002} find that the reaction sequence for CO destruction
suggested by \citet{Prinn:1977} cannot explain the disequilibrium abundance of
CO in Jupiter's troposphere.  A faster reaction pathway for the reduction of
CO to $\rm CH_4$ must be sought.

\citet{Yung:1988} investigate an alternative three-step process involving the
methoxy radical $\rm CH_3O$.  They argue this pathway is more kinetically
favorable because the strong carbonyl ($\rm C = O$) bond in $\rm H_2CO$ is
broken in a separate reaction step from the formation of the four C--H bonds in
methane.  \citet{Yung:1988} propose the following three-stage reaction set:
\begin{equation}
\rm H\, + \, H_2CO \, + \, M \rightarrow CH_3O \, + \, M
\label{reaction:Yung1}
\end{equation}
\begin{equation}
\rm CH_3O \, + \, H_2 \rightarrow CH_3OH \, + \, H
\label{reaction:Yung2}
\end{equation}
\begin{equation}
\rm CH_3OH \, + \, H \rightarrow CH_3 \, + \, H_2O
\label{reaction:Yung3}
\end{equation}

In this sequence, the first reaction, in which the $\rm C = O$ bond is broken,
is the rate-limiting step.  The net reaction resulting from summing reactions
\ref{reaction:Yung1}, \ref{reaction:Yung2}, and \ref{reaction:Yung3} is 
\begin{equation}
\label{reaction:Yung_net}
\rm H_2CO \, + \, H_2 \, + \, H \rightarrow CH_3 \, + \, H_2O.
\end{equation}
We derive the number density of $\rm H_2CO$ by assuming equilibrium with $\rm
H_2$ and CO: $\rm H_2 \, + \, CO \leftrightarrow H_2CO$, which is fast
compared to reaction \ref{reaction:Yung1}.  We note also that the production
of methane ($\rm CH_4$) from $\rm CH_3$ occurs with nearly 100\% efficiency
\citep{Yung:1988}.

The rate of reaction \ref{reaction:Yung1} has never been measured in the
laboratory, but the rate of the reverse reaction is known
\citep{Page:1989}:
\begin{equation}
\label{equation:Yung1R_rate0}
\rm \mathnormal{k}^0_{\rm \ref{reaction:Yung1},\:R} = 1.4\times
10^{-6}\;\left(\frac{\mathnormal{T}}{K}\right)^{-1.2}
\,exp\left(-\frac{7,800\:K}{\mathnormal{T}}\right)\;cm^3\;s^{-1}.
\end{equation}
Equation \ref{equation:Yung1R_rate0} represents the low-pressure limit of
the rate of the reverse of reaction \ref{reaction:Yung1} in the temperature
range relevant for this planet.  For the high-pressure limit, we adopt
equation (11) from \citet{Bezard:2002}:
\begin{equation}
\label{equation:Yung1R_rate_inf}
\rm \mathnormal{k}^{\infty}_{\rm \ref{reaction:Yung1},\:R} = 1.5\times
10^{11}\;\left(\frac{\mathnormal{T}}{K}\right)\cdot 
exp\left(-\frac{12,880\:K}{\mathnormal{T}}\right)\;s^{-1}.
\end{equation}
The net rate constant for the reverse reaction \ref{reaction:Yung1},R can be
expressed as
\begin{equation}
\label{equation:Yung1R_rate}
\rm \mathnormal{k}_{\rm \ref{reaction:Yung1},\:R} = 
\frac{\mathnormal{k}^0_{\rm \ref{reaction:Yung1},\:R}
\mathnormal{k}^{\infty}_{\rm \ref{reaction:Yung1},\:R} } {
\mathnormal{k}^0_{\rm \ref{reaction:Yung1},\:R}\,
\mathnormal{n} \, + \, \mathnormal{k}^{\infty}_{\rm \ref{reaction:Yung1},\:R}
},
\end{equation}
where $n$ [$\rm cm^3\;s^{-1}$] here is the number density of molecules in the
atmosphere \citep{Bezard:2002}.  It is related to pressure through the ideal
gas equation of state.

Using the proper eddy mixing timescale for chemical quenching derived by
\citet{Smith:1998}, \citet{Griffith:1999} and \citet{Bezard:2002} demonstrate
that the kinetic rates derived from the \citet{Yung:1988} reactions fit
observations of CO in disequilibrium on the brown dwarf Gl 229b and on
Jupiter.  As noted above, the reaction proposed by \citet{Prinn:1977} is too
slow to account for the observed concentrations of CO in these objects.  We
therefore follow \citet{Griffith:1999} and \citet{Bezard:2002} in this
treatment and assume that the catalyzed \citet{Yung:1988} reactions
\ref{reaction:Yung1}--\ref{reaction:Yung3} are the most efficient mechanism
for the conversion of CO to $\rm CH_4$.

For the purpose of determining the CO quench level on HD 209458b, we use the
measured reaction rates (eq. \ref{equation:Yung1R_rate0} --
\ref{equation:Yung1R_rate}) along with thermochemical data to determine
$\tau_{\rm chem}$.  This quantity, which is a strong function of both pressure
and temperature, represents the timescale over which CO can remain in
disequilibrium \citep{Prinn:1977}.  Conservation of mass (eq.
\ref{equation:c1}) dictates that the rate of CO destruction must equal the
rate of $\rm CH_4$ production.  So, $\tau_{\rm chem}$ also represents the
timescale for conversion of $\rm CH_4$ back to CO.  This would not be so if
other molecules competed effectively for the available carbon atoms.  At these
pressures and temperatures, though, CO and $\rm CH_4$ are the predominant
carbon-bearing species (see \S \ref{section:Equilibrium_Considerations}).
Hereafter, we simply refer to $\tau_{\rm chem}$ as the CO/$\rm CH_4$
interconversion timescale.

The CO/$\rm CH_4$ interconversion timescale is given by \citet{Bezard:2002}:
\begin{equation}
\label{equation:tau_chem1}
\tau_{\rm chem} = \frac{X_{\rm CO}}{-\mathrm{d}X_{\rm CO}/\mathrm{dt}} 
= \frac{X_{\rm CO}}{\mathnormal{n^2\,k_{\rm \ref{reaction:Yung1}}}\cdot X_{\rm H}X_{\rm H_2CO}},
\end{equation}
where $k_{\rm \ref{reaction:Yung1}}$ is the reaction rate of \ref{reaction:Yung1}.  An extra
factor of $n$ arises in the denominator from writing the expression in terms
of CO, H, and $\rm H_2CO$ mole fractions rather than concentrations.  

As has been noted, however, only the rate of the reverse of reaction
\ref{reaction:Yung1} has been measured.  We therefore use the equilibrium
constant of the chemical reaction 
\begin{equation}
\label{reaction:CH3O_equil}
\rm CH_3O \, \leftrightarrow \, CO \, + \, 3/2\;H_2
\end{equation}
to express $\tau_{\rm chem}$ in terms of equation \ref{equation:Yung1R_rate}.
The equilibrium constant of reaction
\ref{reaction:CH3O_equil} is
\begin{equation}
\label{equation:K_CH3O} \rm \mathnormal{K}_{\ref{reaction:CH3O_equil},\,\rm eq} =
exp\left[-\frac{\Delta_f\mathnormal{G}(CO) - \Delta_f\mathnormal{G}(CH_3O)}{\mathnormal{RT}}\right].
\end{equation}
The Gibbs free energy of formation of $\rm CH_3O$ cannot be found in the JANAF
tables.  We use values for $\rm \Delta_f\mathnormal{G}(CH_3O)$ from
\citet{Tsang:1986}.  

Using these thermochemical data along with equation
\ref{equation:Yung1R_rate}, we can express $\tau_{\rm chem}$ in a form that
can be evaluated numerically:
\begin{equation}
\label{equation:tau_chem2} 
\rm \tau_{chem}(\mathnormal{p}\rm,T) =
\frac{\mathnormal{X}_{CO}}{\mathnormal{n}\,\mathnormal{k}_{\ref{reaction:Yung1},\:R}
\cdot \mathnormal{X}_{CH_3O}} = 
\frac{\mathnormal{K}_{\ref{reaction:CH3O_equil},\,\rm eq}\cdot \mathnormal{k}_b\mathnormal{T}}
{\mathnormal{k}_{\ref{reaction:Yung1},\:R}\,\mathnormal{X}_{\rm H_2}^{3/2} 
\mathnormal{p}^{5/2}},
\end{equation}
where $k_{\rm b}$ is Boltzmann's constant.  The first equality in equation
\ref{equation:tau_chem2} is analogous to equation \ref{equation:tau_chem1}
but for reaction \ref{reaction:Yung1} in the \em reverse \em direction.  All
intermediate reactions in the system are assumed to be in equilibrium.  The
second equality results from substituting equation \ref{equation:K_CH3O} in
for the ratio $X_{\rm CO}/X_{\rm CH_3O}$.  

We show values for $\tau_{\rm chem}$ in Table \ref{table:tau_chem} at the same
pressures and temperatures chosen in Table \ref{table:CO_fraction}.  Table
\ref{table:tau_chem} demonstrates that for $p < 1$ bar, $\tau_{\rm chem} \gg
10^5\;\rm s$ for all $\rm T \leq 2000\;K$, which is an upper limit to the
temperature at HD 209458b's photosphere.  This result immediately rules out
horizontal quenching as an important disequilibrium process at the
photosphere.  The timescale for horizontal mixing $\tau_{\rm dyn,\,h}\sim
10^5\;s$, so that $\tau_{\rm chem} \gg \tau_{\rm dyn,\,h}$ on both the day and
night sides of the planet.

Deeper down ($p > 10$ bar), CO/$\rm CH_4$ interconversion is fast ($\tau_{\rm
chem} \ll 10^5$ s).  At this pressure, the chemical interconversion timescale
is less than both the horizontal and vertical mixing times.  We would
therefore expect chemical equilibrium of CO/$\rm CH_4$ to hold in the
atmosphere at $p \gtrsim 10$ bar.  The top of this region ($p \sim 10$ bar) is
the approximate ``quench level.''  Note that the precise altitude of quenching
also depends on atmospheric temperature and wind speeds, which are themselves
highly variable \citep{Cooper:2005}.  As a consequence, the quench level is
not an exact isobar.  Referencing Table \ref{table:CO_fraction}, we see that
abundant CO at the photosphere of this planet is likely because vertical
transport processes are probably efficient \citep{Showman:2002}, and the
chemical equilibrium CO concentration at the quench level is likely to be
quite high.

We now discuss a coupled atmospheric dynamics and disequilibrium CO/$\rm CH_4$
chemistry model to predict CO concentrations for a variety of atmospheric
conditions.  Our simulation results (see \S \ref{section:Simulation_Results})
verify the general validity of these considerations.

\section{Coupled Chemistry/Dynamics Model}
\label{section:Model}

\subsection{The Atmospheric Dynamics Model}
\label{subsection:Dynamics_Model}

We adapt the model of \citet{Cooper:2005} to include simple carbon chemistry
based on the discussions in \S \ref{section:Equilibrium_Considerations} and
\S \ref{section:Chemical_Timescales}.  Our model employs the ARIES/GEOS
Dynamical Core, version 2 \citep[AGDC2;][]{Suarez:1995}.  The AGDC2 solves the
primitive equations of dynamical meteorology, which are the foundation of
numerous climate and numerical weather prediction models \citep{Holton:1992,
Kalnay:2003}.

The primitive equations are a simplification of the full Navier-Stokes
equations of fluid mechanics, which assume hydrostatic balance of each
vertical column of atmosphere.  This is equivalent to assuming that the
vertical acceleration is negligible compared to the buoyancy.  The
approximation holds in stably stratified regions for ``shallow'' flow, in
which the vertical depth is much smaller than the horizontal extent
\citep[][Ch. 2]{Kalnay:2003}.  Heavy irradiation from the star extends the
radiative zone of this planet down to $p \sim 1$ kbar.  This is $\sim\,$1/10
of the radius \citep{Burrows:2003, Chabrier:2004, Iro:2005}, leading to a
horizontal to vertical aspect ratio of about 10:1. 

The AGDC2 is a finite-difference model, which discretizes the dynamical
variables onto a staggered latitude-longitude C-grid \citep{Arakawa:1977}.
The AGDC2 consists of a set of subroutines that compute the time tendencies of
the prognostic variables, including the zonal and meridional winds, potential
temperature \citep[][p. 52]{Holton:1992}, surface pressure, and an arbitrary
number of passive tracers.  At each time step, the AGDC2 simply updates the
time tendencies to include the effects of the dynamics.  All physical
tendencies (e.g., chemistry, radiation, etc.), as well as temporal and spatial
filtering needed to control computational instabilities, are performed outside
of the AGDC2.

It is not possible in general to find numerically stable solutions to the
primitive equations without viscosity \citep{Dowling:1998}.  The standard
approach, which we adopt here, is to add diffusive terms to the momentum,
energy conservation, and tracers equations.  We have experimented with the
diffusion parameters, espousing the philosophy of \citet{Polvani:2004} to use
as weak hyperdiffusion as possible consistent with numerically stable
solutions.  For the simulations discussed here, we use $4^{\rm th}$-order
hyperdiffusion, with coefficients set to damp grid-scale instabilities over a
timescale of $\sim\!30\;\rm minutes$.

For all simulations, we take the "top" layer of the model to be at 1 mbar.
Although a proper model of the upper atmosphere would require the model top be
placed at a lower pressure, we generally ignore the upper atmosphere in this
work.  Modeling the high atmosphere would require a proper treatment of
stratospheric photochemistry and the possible effects of non-local
thermodynamic equilibrium in the radiation field at sub-millibar pressures
\citep{Liang:2004, Barman:2002}.  The atmosphere spans $\sim\!15$ pressure
scale heights between our input top pressure (1 mb) and the planet's
radiative-convective boundary (1 kbar).  By comparison, the pressure in the
Earth's atmosphere spans only 2--3 pressure scale heights from the surface to
the tropopause.  We use 40 layers evenly spaced in log pressure.

We assume in this treatment that synchronous rotation holds in the convective
interior.  The consequences of this assumption are twofold: (1) the rotational
period is fixed to the 3.5 day orbital period (leading to a moderately strong
Coriolis force), and (2) heating by the star occurs on one hemisphere of the
planet continually.  This assumption is valid for many hot Jupiters (such as
HD 209458b), which experience rapid tidal circularization of their orbits
\citep{Guillot:1996}.  Atmospheric tides may cause the orbit to be slightly
asynchronous, as noted by \citet{Showman:2002}.  But large deviations from
synchronous rotation are unlikely on HD 209458b without the presence of
another (thus far unseen) planet \citep{Laughlin:2005b}.  We discuss in \S
\ref{subsection:Radiation} that the asynchronous component (a factor of 2 in
the extreme case, though more likely in the 10--20\% range) will not greatly
affect the outcome of our predictions for $\rm CO/CH_4$ chemistry (see \S
\ref{section:Discussion}).  But it would be interesting in future studies to
explore the dynamics of hot Jupiters under the alternate assumption of
variable heating due to non-zero eccentricity, large obliquity, or strong
thermal tides.

We follow \citet{Cooper:2005} for the scalar input parameters, ignoring small
variations in the gravitational acceleration, heat capacity, and mean
molecular weight in the atmosphere.  We set these values to constants: g =
$\rm 9.42\;m\;s^{-2}$, $\rm \mu = 1.81\times 10^{-3}\;kg\;mol^{-1}$, and
$c_{\rm p} = \rm 1.43 \times 10^4\;J\;kg^{-1}\;K^{-1}$.  The values of these
constants have been chosen to be intermediate between the values they would
have at 1 mbar and 3 kbar because $\rm H_2$ dissociation is expected to occur
in the deepest layers of the model.  For the chemistry, we assume hydrogen is
molecular everywhere, which holds in the upper layers of interest here (see
\S \ref{section:Chemical_Timescales}).

\subsection{Coupling $\rm CO/CH_4$ Chemistry}
\label{subsection:Model_CO_Chemistry}

The AGDC2 includes horizontal advection for an arbitrary number of passive
tracers.  We denote these as $q_k$, with $k = 1, 2, ...,km$.  As few as
possible is desirable, as the tracer equations add appreciably to the total
computations required per time step.  

Tracers are passive in the sense that they are advected by the flow but do not
themselves influence the dynamics.  Unless interdependencies are added, the
tracer solutions in the AGDC2 do not affect the other prognostic variables (or
each other).  The equations operating on each $q_k$ are of the same form as
those used for potential temperature advection \citep{Suarez:1995}.  The
advection scheme is fourth-order in this version of the dynamical core.  

We add CO as a passive tracer to the AGDC2.  The other carbon and oxygen
bearing species, including $\rm CH_4$, $\rm H_2O$, $\rm CH_3O$, $\rm HCO$,
etc., are either very short-lived or indirectly obtainable through mass
balance (eq. \ref{equation:c1} and \ref{equation:c2}).  To confirm this,
we calculated lifetimes for the intermediate species in the \citet{Yung:1988}
sequence.  These were all much shorter than our $\rm \tau_{chem}$, which
describes the timescale for interconversion of CO and $\rm CH_4$.  For the
purposes of this investigation, it is sufficient therefore to consider just a
single tracer.

We follow equation (2) of \citet{Smith:1998} to express the effect of the
dynamics on $\rm CO/CH_4$ chemistry in terms of the quantities we know,
$X_{\rm CO,\:eq}$ and $\tau_{\rm chem}$.  The idea is to relax $X_{\rm CO}$ to
its value in chemical equilibrium, $X_{\rm CO,\:eq}$:
\begin{equation}
\label{equation:Relax_X_CO}
\frac{DX_{\rm CO}}{Dt} = -\frac{X_{\rm CO} - X_{\rm CO,\:eq}}{\tau_{\rm chem}}.  
\end{equation}
We write $D/Dt$ (not $d/dt$ or $\partial/\partial t$) to emphasize we mean the
total (or material) derivative, \emph{not} the local derivative.  As usual,
the advection terms must be included when the equations are written in the
Eulerian frame \citep[][p. 28--31]{Holton:1992}.

In equation \ref{equation:Relax_X_CO}, we see that if $\tau_{\rm chem}$ is
short relative to dynamical timescales, chemical equilibrium of CO will be
attained in the atmosphere.  In regions where $\tau_{\rm chem}$ is longer than
$\tau_{\rm dyn}$, CO will not be in equilibrium.  In the former case, the CO
abundance will depend solely on the local temperature and pressure (which
themselves vary greatly throughout the region of integration).  In the latter
situation, whether the actual value of $X_{\rm CO}$ is greater than or less
than the equilibrium value depends on global atmospheric conditions.  

In the troposphere of Jupiter, for example, $X_{\rm CO}$ is extremely high
relative to $X_{\rm CO,\:eq}$ because convection mixes CO upwards from the
quench level (at several hundred bars), where the equilibrium abundance of CO
is appreciable \citep{Bezard:2002}.  The situation of HD 209458b is different;
the atmosphere in the 10 mbar to 10 bar region is stable to turbulent
convection due to the huge flux of stellar radiation from above
\citep{Guillot:1996}.  As discussed by \citet{Showman:2002}, however, vertical
velocities of $\sim\!20\rm\;m\;s^{-1}$ are plausible on this planet, implying
vigorous vertical transport.  The CO concentration for $p < 10\;\rm bar$
depends critically on the vertical flow, which we will discuss in
\S \ref{section:Analysis_Interpretation}.

\subsection{Newtonian Heating Scheme}
\label{subsection:Newtonian_Scheme}

We defer to future work the ambitious task of coupling atmospheric
dynamics to a true radiative transfer code.  Rather, we approximate the
effects of radiation using a simple Newtonian heating prescription.  In this
scheme, the thermodynamic heating rate at each grid point is proportional to
the difference between the prescribed 3D radiative-equilibrium temperature and
the true local temperature.

As in \citet{Cooper:2005}, we start with the 1D radiative-equilibrium $T(p)$
profile of \citet{Iro:2005}.  Their radiative-equilibrium profile, which we
call $T_{\rm Iro}(p)$, assumes heat redistribution over the entire globe.  The
calculation includes the major sources of gaseous opacity: Rayleigh
scattering, collision induced absorption, bound-free and free-free absorption,
molecular rovibrational bands from $\rm H_2O$, CO, $\rm CH_4$, and TiO, and
resonance lines from the alkali metals.  \citet{Iro:2005} calculated a
line-by-line solution to the monochromatic radiative transfer equation using
the \citet{Goukenleuque:2000} model.  This code makes the two-stream
approximation \citep[see e.g.,][]{Chamberlain:1987}, and the atmosphere is
taken to be plane-parallel.  We also use values published by \citet{Iro:2005}
for $\tau_{\rm rad}(\mathnormal{p})$, which is the radiative relaxation time
constant.  This is an increasing function of pressure.  The value of
$\tau_{\rm rad}$ is $\sim\!3$ hours at 1 mbar (which is the topmost layer of
our model) and increases to $\sim\!1$ year near 10 bar.

We use the algorithm described in \citet{Cooper:2005} to calculate $T_{\rm
night}(p)$ and $T_{\rm ss}(p)$ from $T_{\rm Iro}(p)$.  These are the
radiative-equilibrium profiles on the nightside, which receives no direct
heating, and at the substellar point, where the star is directly overhead.  We
assume in these calculations that nightside radiative-equilibrium temperatures
depend only on pressure, not longitude and latitude.  In this procedure, we
prescribe a temperature difference in radiative equilibrium between the
substellar point and the nightside, $\Delta T_{\rm eq}(p) = T_{\rm ss}(p) -
T_{\rm night}(p)$.  This is an adjustable parameter, which governs the
strength of the radiative forcing.  Its has been chosen to be logarithmic
in pressure, so the profiles converge deep in the atmosphere.

The substellar and nightside radiative-equilibrium profiles are then computed
by balancing the total flux radiating to space at the top of the atmosphere.
The nightside and integrated dayside contributions must sum to the known
globally averaged value, $4\pi R_{\rm p}^2\sigma T_{\rm Iro}^4$:
\begin{equation}
\label{equation:flux_balance}
4\pi R_{\rm p}^2\sigma T_{\rm Iro}^4 = 2\pi R_{\rm p}^2\sigma T_{\rm night}^4 + R_{\rm p}^2 
\int\!\!\!\!\int_{\rm day}
\sigma T_{\rm day}^4(\lambda,\phi)\,\rm \cos\phi\,d\phi\,d\lambda.
\end{equation}
The coordinates ($\lambda, \phi$) are longitude and latitude, respectively.

The flux balance in equation \ref{equation:flux_balance} yields a
transcendental equation for $T_{\rm night}$ at the top of the atmosphere in
terms of the free parameter, $\Delta T_{\rm eq}$:
\begin{equation}
\label{equation:T_night_top} 
4T_{\rm iro}^4 = 3T_{\rm night}^4 + (T_{\rm night} + \Delta T_{\rm eq})^4.
\end{equation}
In equation \ref{equation:T_night_top}, 2 factors of $T_{\rm night}^4$ come
from the nightside; the integral over the dayside contributes 1 more factor
of $T^4_{\rm night}$, as well as the term involving $\Delta T_{\rm eq}$.
Equation \ref{equation:T_night_top} can be solved numerically; we use
Newton-Raphson \citep[see][]{Press:1992}.  $T_{\rm night} = 500\;\rm K$ is the
solution for $\Delta T_{\rm eq} = 1000\;\mathrm{K}$ and $T_{\rm Iro} =
1000\rm\;K$.

The above radiation balance is only computed once---at the top model
layer---to ensure the outgoing fluxes are self-consistent.  We emphasize this
is not a rigorous treatment of radiation.  The goal is simply to use
\citet{Iro:2005}'s radiative-equilibrium profile, which is one-dimensional, to
prescribe the Newtonian heating rate over the model grid in a way that does
not significantly alter the radiation budget of the atmosphere.  

\begin{figure}
\includegraphics[scale=0.400, angle=-90]{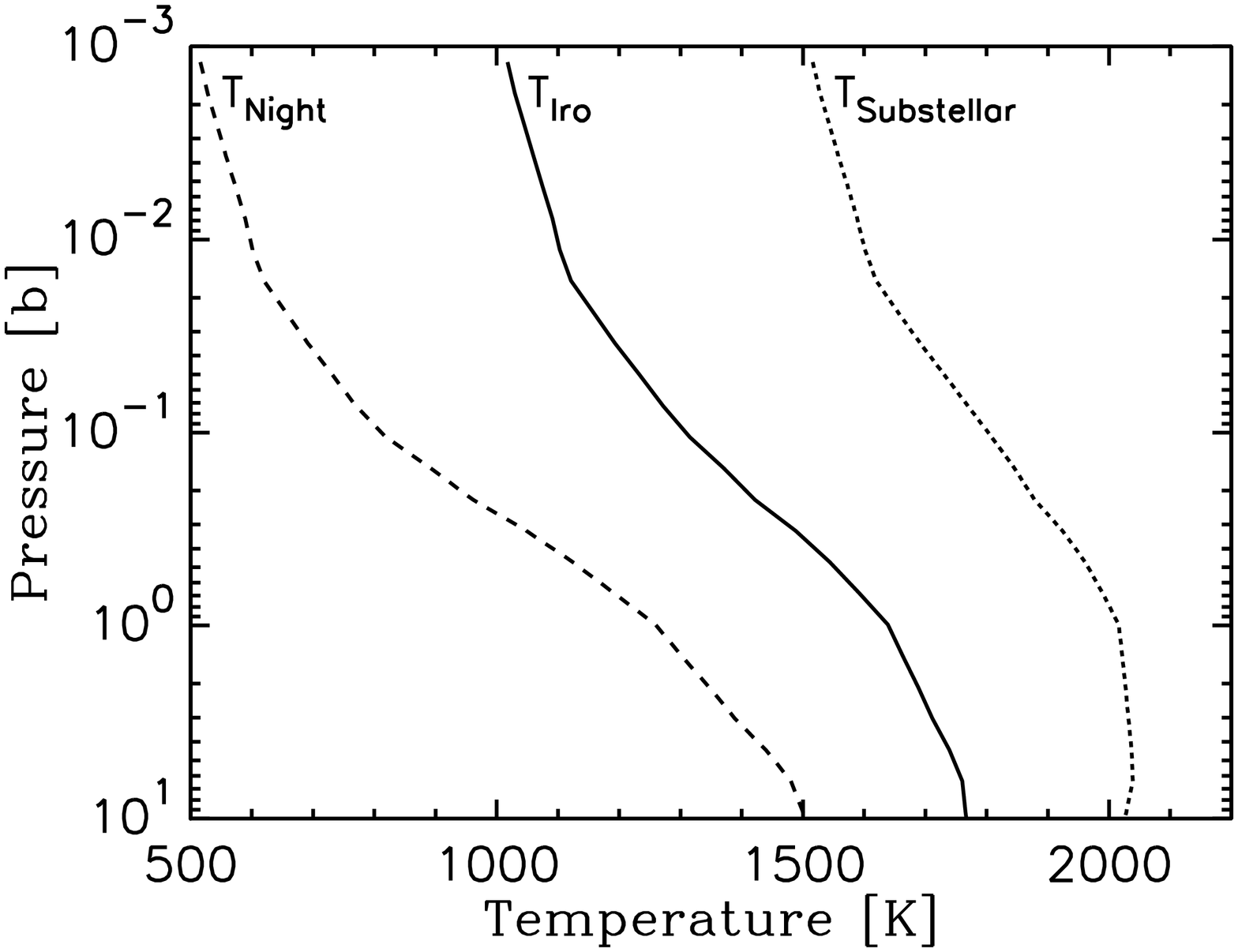}
\caption{Radiative-equilibrium $T(p)$ profiles used in the Newtonian
heating scheme described in \S \ref{subsection:Newtonian_Scheme}.  The
profiles shown are the nightside radiative-equilibrium profile (long-dashed
line), \citet{Iro:2005}'s globally averaged radiative-equilibrium profile
(solid line), and the radiative-equilibrium profile at the substellar point
(short-dashed line).  At the topmost layer of the model atmosphere (1 mbar),
we have set the temperature difference between the substellar point and the
nightside to 1000 K.  The nightside and substellar profiles are balanced so
the integrated outgoing flux at the top of the atmosphere equals $4\pi R_{\rm
p}^2\sigma T_{\rm Iro}^4$.  The profiles converge logarithmically with
increasing pressure (depth).
}
\label{figure:rad_eq_profiles}
\end{figure}

We show in Figure \ref{figure:rad_eq_profiles} the radiative-equilibrium
profiles generated by this method from pressures of 1 mbar to 10 bars.  We
plot $T_{\rm Iro}(p)$ as well for comparison.  The profiles plotted are for
our nominal HD 209458b model, in which we have set $\Delta T_{\rm eq} = T_{\rm
ss}-T_{\rm night} = 1000\;\rm K$ at the top of the atmosphere (see
\S \ref{section:Simulation_Results} for a more complete description).  We only
show the profiles down to the 10 bar level because the radiative time
constants deeper than that are very long ($\tau_{\rm rad} \rm \gg 10^7\;s$;
see \citet{Iro:2005}).  As in \citet{Cooper:2005}, we have simply turned off
heating for all pressures greater than 10 bar.

\subsection{Limb Effects}
\label{subsection:Limb_Effects}

Two limb effects may be important at low pressures.  We have not attempted
here to model them in detail.  But to account for them, we use a modified form
of \citet{Cooper:2005}, equation(2).

First, near the limb, the angle of incidence of the impinging radiation is
large.  The optical path length is longer in the slant geometry than at normal
incidence.  Hence, the stellar radiation will not penetrate as deeply near the
limb but will be absorbed higher up in the atmosphere.  This causes additional
heating at pressures significantly less than 1 bar.  Second, the star is not a
point source; it has finite angular size in the planet's sky.  This creates a
small region of continuous heating past the day-night terminator, which we
define here as $\rm 90\degr$ from the substellar point.  Here, the disk of the
star is partly visible.  

\begin{figure}
\includegraphics[scale=0.470]{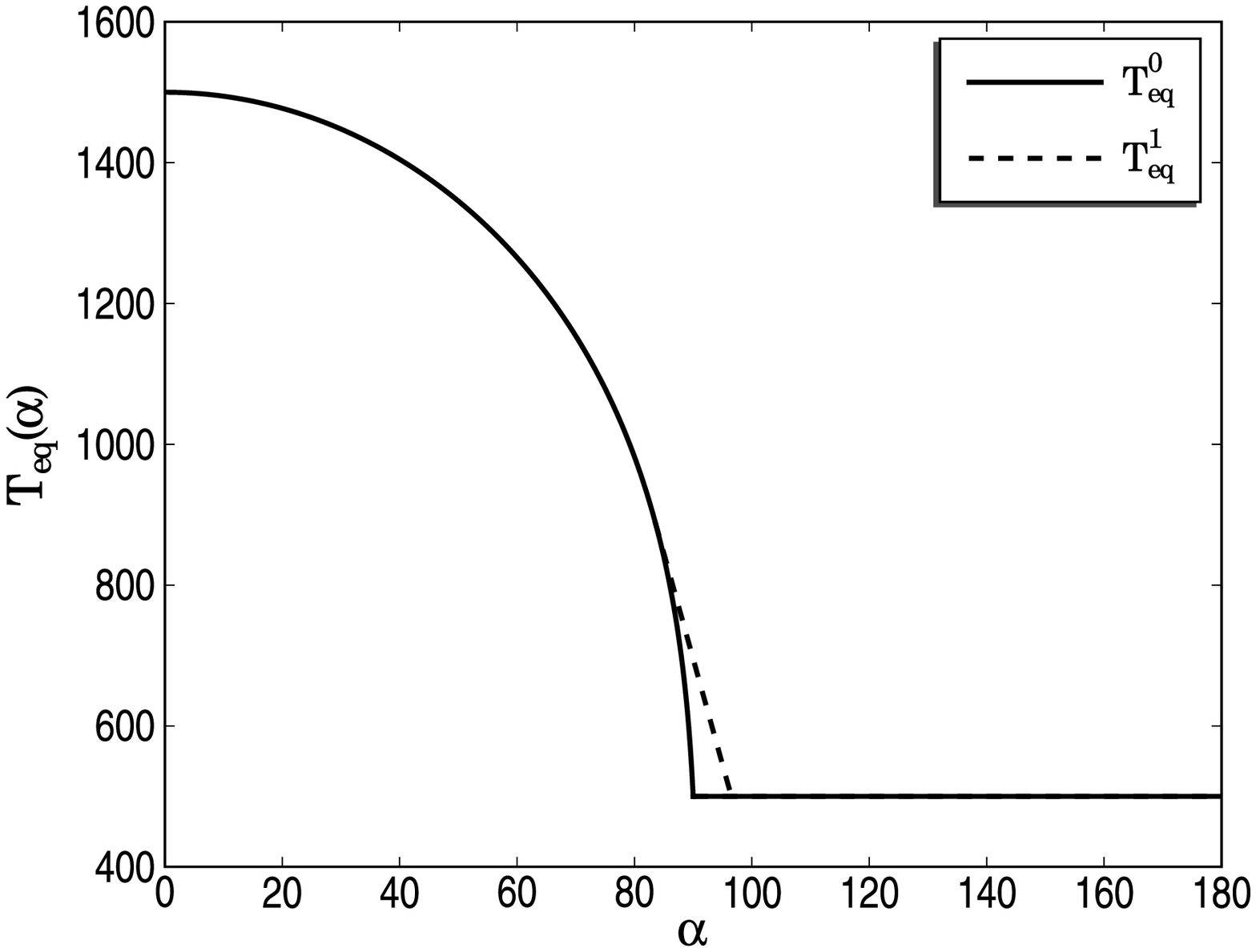}
\caption{The angular dependence of the 3D radiative-equilibrium profile, which
is symmetric about the star-planet line.  We define $\alpha$ to be the angle
between the local normal vector at a particular longitude and latitude
($\lambda$, $\phi$) and a ray from the planet's center passing through the
substellar point $(\lambda = 0\degr$, $\phi = 0\degr)$.  This is a slightly
warmer profile near the limb than that used by \citep{Cooper:2005}.
}
\label{figure:T_eq_alpha}
\end{figure}

These considerations are potentially significant for tracking the CO
distribution near the photosphere because the limb is the region directly
probed by transmission spectroscopy.  We attempt to account for this by
modifying \citet{Cooper:2005}'s equation (2).  The new profile, shown in
Figure \ref{figure:T_eq_alpha}, is identical to that of \citet{Cooper:2005}
over most of the day and night sides.  But it has a smoother variation of
temperature across the terminator, which remains fairly hot $90\degr$ away
from the substellar point (where half the stellar disk is still visible).  It
also increases the total radiative energy in the radiative-equilibrium system,
which is realistic owing to the effects described above.

\section{Simulation Results}
\label{section:Simulation_Results}

For this study, we have performed a suite of simulations with different input
parameters, which all support the general conclusions of this work
(\S \ref{section:Implications}).  For most of this discussion, we focus on the
results of two simulations in particular.  We call them the ``nominal'' and
``cold'' models, respectively.  Other simulations are alluded to in this
section and in
\S \ref{section:Analysis_Interpretation}--\ref{section:Implications}.

The nominal model contains input parameters identical to those used by
\citet{Cooper:2005}, which are tuned for HD 209458b specifically (although the
radiative-equilibrium profile has been slightly adjusted to account for limb
effects; see \S \ref{subsection:Limb_Effects}).  For the nominal model, we
have used 1000 K for the radiative-equilibrium temperature difference between
the substellar point and the nightside (at the top of the atmosphere, as
described in \S \ref{subsection:Newtonian_Scheme}).  The cold model is a
parameter variation on the nominal model.  It is representative of an
imaginary EGP that is cooler than HD 209458b.  The radiative-equilibrium
temperatures for the cold model are similar to those expected for the
transiting planet TrES-1, which orbits a K-type star \citep{Fortney:2005a,
Barman:2005}. 

We use for the cold model a new radiative-equilibrium profile, $T_{\rm
Cold}(p)$.  This profile is 300 K cooler at every pressure level than the
profile used for the nominal model: $T_{\rm Cold}(p) = T_{\rm Iro}(p) -
300\;K$.  For the cold case, we have also decreased the assumed
radiative-equilibrium temperature difference between the substellar point and
the nightside; for this model, we use $\Delta T_{\rm eq} = 700\rm \;K$.  This
is because we would expect cooler planets to have lower radiative forcing in
general.  With these adjustments, the procedure described in
\S \ref{subsection:Limb_Effects} was rerun on the cold profile to determine the
3D radiative-equilibrium profile for the cold case.  At the topmost layer (1
mbar), the substellar and nightside profiles are $T_{\rm ss} = 1085\rm \;K$
and $T_{\rm night} = 385\rm \;K$, respectively.

We use the same values for the radiative relaxation time constant, $\tau_{\rm
rad}(\mathnormal{p})$, as for the nominal case.  For a much cooler planet than
HD 209458b, $\tau_{\rm rad}(\mathnormal{p})$ would have to be re-evaluated,
since radiative processes are temperature dependent.  But this ``cold'' model
is not enough colder than the nominal case for the radiative heating rates to
differ by more than a factor of $\sim\!2$.  All other input parameters
(acceleration of gravity, heat capacity, mean molecular weight, and radius)
have been kept the same as in the nominal model.  It should be emphasized,
therefore, that the cold model is \emph{not} a model of TrES-1 or any other
known EGP in particular.  It is simply a parameter variation on the nominal
model, which is intended to elucidate the role of atmospheric temperature in
determining the $\rm CO/CH_4$ chemistry.

For all simulation runs presented here, the horizontal grids contain 72 points
in longitude and 45 points in latitude.  We have run the models for $t_{\rm
sim}$ = 1000 Earth days.  This is long enough for the atmosphere model to
equilibrate on layers having $p \la 3$ bar, since the radiative time constants
are fairly short---less than a month---in this region.  We have verified that
the kinetic energy in these low-pressure layers does not increase with time
after 500--1000 days.  We use a fairly small time step for all simulations:
$\rm \Delta \mathnormal{t} = 50\;s$.  The small time step helps to control
grid-scale instabilities.  We discuss these issues further in
\S \ref{section:Discussion}.

We assume zero initial zonal and meridional winds: $(u,\;v)$ = (0, 0).  We
also start the simulations with the temperature on every layer isothermal and
equal to $T_{\rm Iro}$.  We have verified that our solutions for the flow
geometry---in the layers of interest here---hold even in the case of an
initially strong retrograde wind \citep{Cooper:2005}.  For the initial tracer
concentration, we assume chemical equilibrium of CO at $t_{\rm sim} = 0$.  We
will discuss the effects of the initial tracer distribution in
\S \ref{subsection:Initial_Conditions}.

\subsection{Temperature and Winds}
\label{subsection:T_winds}

\begin{figure}
\caption{
Temperature (grayscale) and winds (arrows) for our ``nominal'' model.  This is
a snapshot of the simulation at $t_{\rm sim}$ = 1000 days.  Maximum wind
speeds in each frame are (a) 6600 $\rm m\;s^{-1}$, (b) 4300 $\rm m\;s^{-1}$,
and (c) 3500 $\rm m\;s^{-1}$.  This is the HD 209458b nominal model, which
uses parameters appropriate for the atmosphere of that planet.  
}
\label{figure:T_winds_nominal}
\end{figure}

We focus here on results from the nominal and cold models in the 10--1000 mbar
range.  These are the ``visible layers'' of EGP atmospheres; i.e., the layers
that directly contribute to the outgoing radiation from the planet to space.
The results of our integrations for the nominal model can be seen in Figure
\ref{figure:T_winds_nominal}.  Figure \ref{figure:T_winds_nominal} spans about
5 scale heights in pressure, encompassing the region crucial for observations
of close-in EGPs.  Despite adjustments to the cooling scheme (see \S
\ref{subsection:Limb_Effects}), the temperatures, wind speeds, and flow
geometry of the nominal model are broadly consistent with simulations
published by \citet{Cooper:2005}.  

Strong temperature contrasts on isobars are evident in the simulation for all
three layers shown.  As seen in panel (a) of Figure
\ref{figure:T_winds_nominal}, the atmosphere is nearly in radiative
equilibrium at low pressures, where the radiative time constants are less than
a day.  At higher pressures, panels (b) and (c), the radiative heating is
weaker and large departures from radiative equilibrium can be seen.  A strong
superrotating jet extending from the equator to the mid-latitudes develops in
the flow.  In the middle panel, (b), the hottest regions of the atmosphere are
blown downstream from the substellar point $\rm (\lambda = 0\degr,\;\phi =
0\degr)$ by about $45\degr$.  This layer at 106 mbar is near the planet's
photosphere.  Longitudinal temperature contrasts are apparent near 1 bar, as
shown in panel (c), where the hot region of the atmosphere is shifted downwind
from the substellar point by a full $90\degr$.

Many features of the flow in the cold model (not shown) are quite similar to
the nominal model, though the temperatures in this atmosphere are on
corresponding isobars are notably cooler (by $\rm \sim\!300\;K$).  From a
dynamics standpoint, the models are basically the same, with similar flow
patterns and wind velocities developing in each case.  The temperature
differences between the nominal and cold models are important for determining
the carbon chemistry, however, since the reaction kinetics depend sensitively
on temperature (see \S \ref{section:Chemical_Timescales}).

\subsection{CO Distribution}

\begin{figure}
\caption{Nominal simulation, with inputs appropriate for HD 209458b.  Shows
the fraction of gaseous carbon present as CO in the (incorrect) equilibrium
picture of carbon chemistry.  The other important carbon-bearing species is
$\rm CH_4$.  Together, CO and $\rm CH_4$ comprise the total carbon budget of
the atmosphere \citep{Lodders:2002}.  The top and middle panes show large
gradients in the equilibrium concentration of CO, which correspond to steep
horizontal temperature gradients (see Figure \ref{figure:T_winds_nominal}).
At 990 mbar, however, temperatures are everywhere higher than 1200 K.  On most
of this layer (bottom pane), CO is therefore thermodynamically favored over
$\rm CH_4$.
}
\label{figure:tr_equil_nominal}
\end{figure}

\begin{figure}
\caption{
The fraction of carbon as CO in the nominal model, in which we have coupled
the chemical kinetics to the dynamics.  The simulation results emphasize that
chemical equilibrium does not hold in this region of the atmosphere.  The
simulation predicts a homogeneous distribution of CO, with CO concentrations
in local cool regions exceeding chemical equilibrium values by many orders of
magnitude.
}
\label{figure:tracers_nominal}
\end{figure}

Our result for the distribution of gaseous carbon in HD 209458b's
atmosphere is shown in Figure \ref{figure:tracers_nominal}.  Contrast this
with Figure \ref{figure:tr_equil_nominal}, which shows how carbon would be
distributed under (fictitious) chemical equilibrium conditions.  Grayscale in
these figures represents the percentage of total atmospheric carbon present as
CO.

As shown in panels (a) and (b) of Figure \ref{figure:tr_equil_nominal}, the
assumption of chemical equilibrium implies that vast regions of HD 209458b's
atmosphere are devoid of CO.  In these cool regions, $\rm CH_4$ is strongly
favored thermodynamically.  Steep gradients in the concentration of CO are
possible wherever the temperature varies rapidly about the line of equal $\rm
CO/CH_4$ abundance (refer to Figure 2 of \citet{Lodders:2002}).  By contrast,
panel (c) shows that near the 1 bar level in the atmosphere, CO is the more
thermodynamically stable compound.  This is because near 1 bar, as shown in
Figure \ref{figure:T_winds_nominal}, the temperature is everywhere higher than
1200 K, which is where the equilibrium abundances of CO and $\rm CH_4$ are
equal (at this pressure).

Our model's result, which considers the extremely slow rate of $\rm CO/CH_4$
interconversion in this region of the planet's atmosphere, shows a completely
different view of the carbon chemistry.  The simulation predicts a nearly
homogeneous distribution of CO in all three layers shown.  CO permeates the
atmosphere in this region, being by far the dominant carbon-bearing species
(at the 98-99\% level). 

Our result for the cold model is similar to the nominal case.  The cold
atmosphere simulation also predicts that carbon is homogenized in the upper
atmosphere.  As in the nominal case, carbon is the major carbon-bearing gas in
the 10--1000 mbar range.  Here, though, 20\% of the carbon is $\rm CH_4$ (as
opposed to 1--2\% in the nominal simulation).  The figure reveals that CO is
likely to be highly abundant near the photospheres of EGPs---both on the
dayside and nightside---even on planets like TrES-1 that are significantly
cooler than HD 209458b.

We have also run a ``hot model,'' in which 300 K was added to the
radiative-equilibrium profile of the nominal model: $\rm T_{Hot} = T_{Iro} +
300\;K$.  The circulation pattern seen for the nominal and cold models (Figure
\ref{figure:T_winds_nominal}) develops for the hot model as well, though the
temperatures on each corresponding layer are higher by $\sim\!200$--300 K.
The temperature is high enough in the hot model for CO to be present in high
abundance everywhere in both the equilibrium and disequilibrium pictures.
Therefore, planets significantly warmer than HD 209458b should contain no $\rm
CH_4$ at all near their photospheres.

Steady-state has been reached for $p < 3$ bar.  But the kinetic energy of the
deeper layers in the model still increases with time after 1000 days.  Large
equator-to-pole temperature variations are seen in the bottom three panels
(spanning the 3--10 bar range).  Wind speeds remain high at these pressures.
Although heating has been turned off for pressures exceeding 10 bar, the
atmosphere is not motionless even at 30 bar.  We attribute this to vertical
transport processes that transfer kinetic energy and angular momentum to the
$p > 10\;\rm bar$ region \citep{Cooper:2005}. 

The vertical wind in the model is quite strong ($\omega$ =
0.05--0.1$\rm\;mb\;s^{-1}$ at 3 bar).  The vertical wind shows evidence of
both small-scale ($\sim 10\degr$) and large-scale ($\sim 45\degr$) structure,
with alternating regions of upwelling and downwelling.  The small-scale
features are likely the result of atmospheric waves.  The large-scale
structures correspond to overturning circulations superimposed on the fast
superrotation, although we emphasize the region is not convective (vertical
motion on this planet is physically much more similar to stratospheric
overturning in the Earth's atmosphere).  Upwelling occurs primarily within
$20\degr$ of the equator, whereas prominent downwelling occurs in the
mid-latitudes.  In the 1--1000 mbar region, there is also a longitudinal
dependence of the vertical wind.  In these upper layers of the model, strong
upwelling occurs in hot regions (which is downstream of the substellar point
at 1 bar), whereas downwelling occurs in cool regions.  This longitudinal
structure is present but much weaker in the 1--10 bar region. 

\section{Analysis / Interpretation}
\label{section:Analysis_Interpretation}

Ultimately, the high CO concentrations we see in these simulations are the
result of chemical quenching.  In general, the distribution of carbon-bearing
species will depend on (1) where in the atmosphere quenching occurs (the
``quench level''), and (2) the chemical equilibrium abundances of CO and $\rm
CH_4$ at the quench level.  The strong vertical wind is capable of
transporting parcels through a pressure scale height ($\sim$ 500 km) in $\sim
10^5\;\rm s$, which is comparable to the timescale for horizontal advection of
parcels across a hemisphere.  This supports the notion that quenching occurs
primarily in the vertical, not the horizontal, due to uniformly long chemical
timescales in the region of disequilibrium (see
\S \ref{section:Chemical_Timescales}).  

The quench level of this model---which demarcates the region where the
equilibrium and disequilibrium views give the same answer---is in the $p$ =
3--5 bar range.  This is of course not exact because quenching depends on
temperature, and considerable equator-to-pole temperature variations exist at
depth in these models \citep{Cooper:2005}.  For example, comparison of the
bottom panels of Figures \ref{figure:tr_equil_nominal} and
\ref{figure:tracers_nominal} reveals that at the equator, quenching likely
occurs closer to $p = 2$ bar.  But at higher latitudes, where the temperature
is lower on these layers, the quench level is deeper ($p = 6$ bar).

The cause of such high CO abundances from 10--100 mbar becomes clear.  The
value of $X_{\rm CO}$ in equilibrium is quite high in the region of quenching.
It is in fact generally higher near 3 bar than it is in deeper layers (not
shown).  This is due to the thermodynamic stability of $\rm CH_4$ at high
pressures, as shown in Table \ref{table:CO_fraction}.  The results clearly
indicate, though, that the quench level (not deeper layers) determines the
photospheric value.  This is due to vertical transport.  The strongest
vertical wind is near the equator, where the temperature is also the highest
(on a given vertical layer).  It is for this reason that the disequilibrium
values of CO in the 1--1000 mbar region are so uniformly high: vertical mixing
is maximum at the equator at the quench level, where temperature and pressure
conditions strongly favor the formation of CO.

We now investigate a simpler view of quenching, which yields very similar
results to the numerical simulations.

\subsection{Eddy Diffusion}
\label{subsection:Eddy_Mixing}

\begin{figure}
{\centering
\includegraphics[scale=0.45]{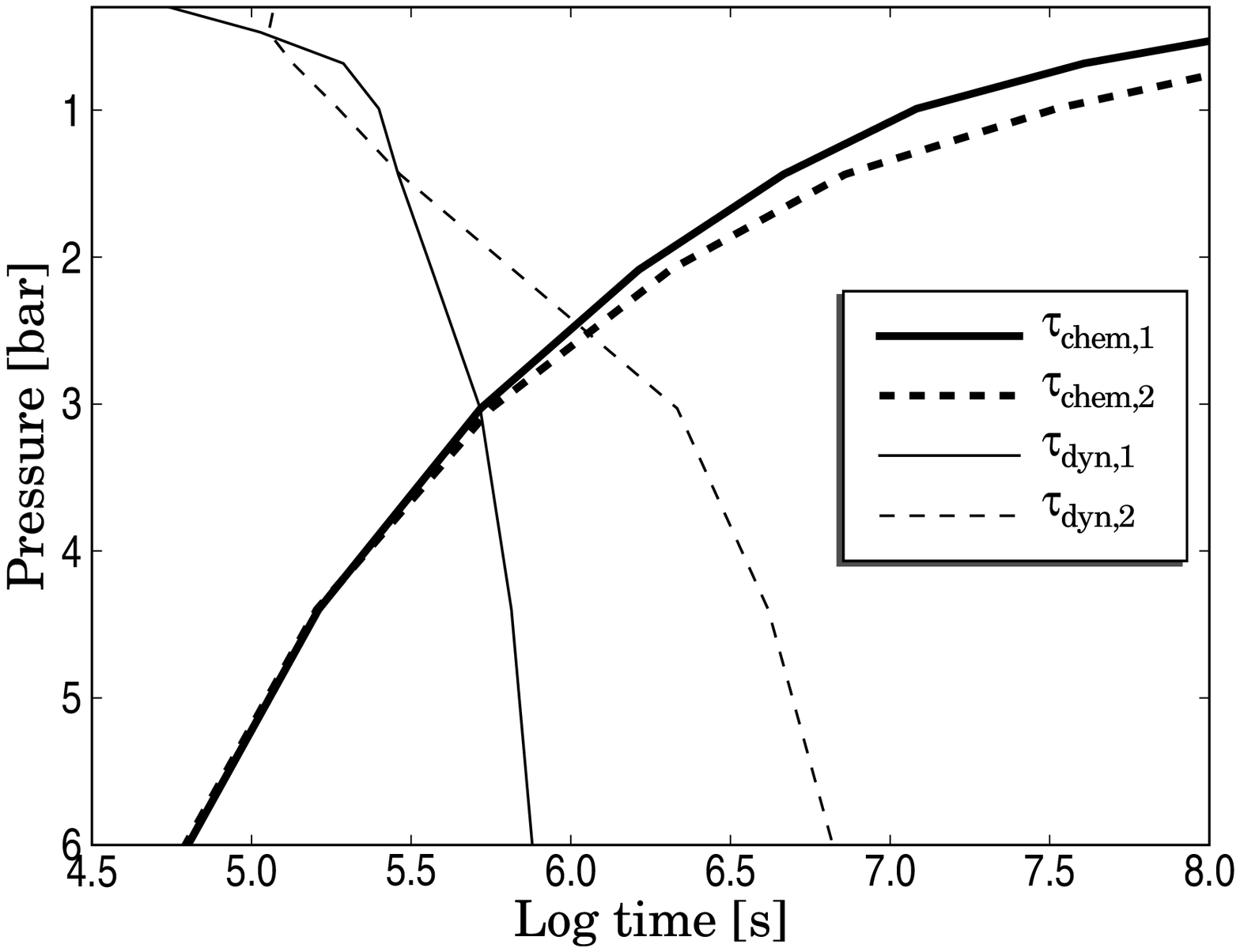}\\[5mm]
\includegraphics[scale=0.45]{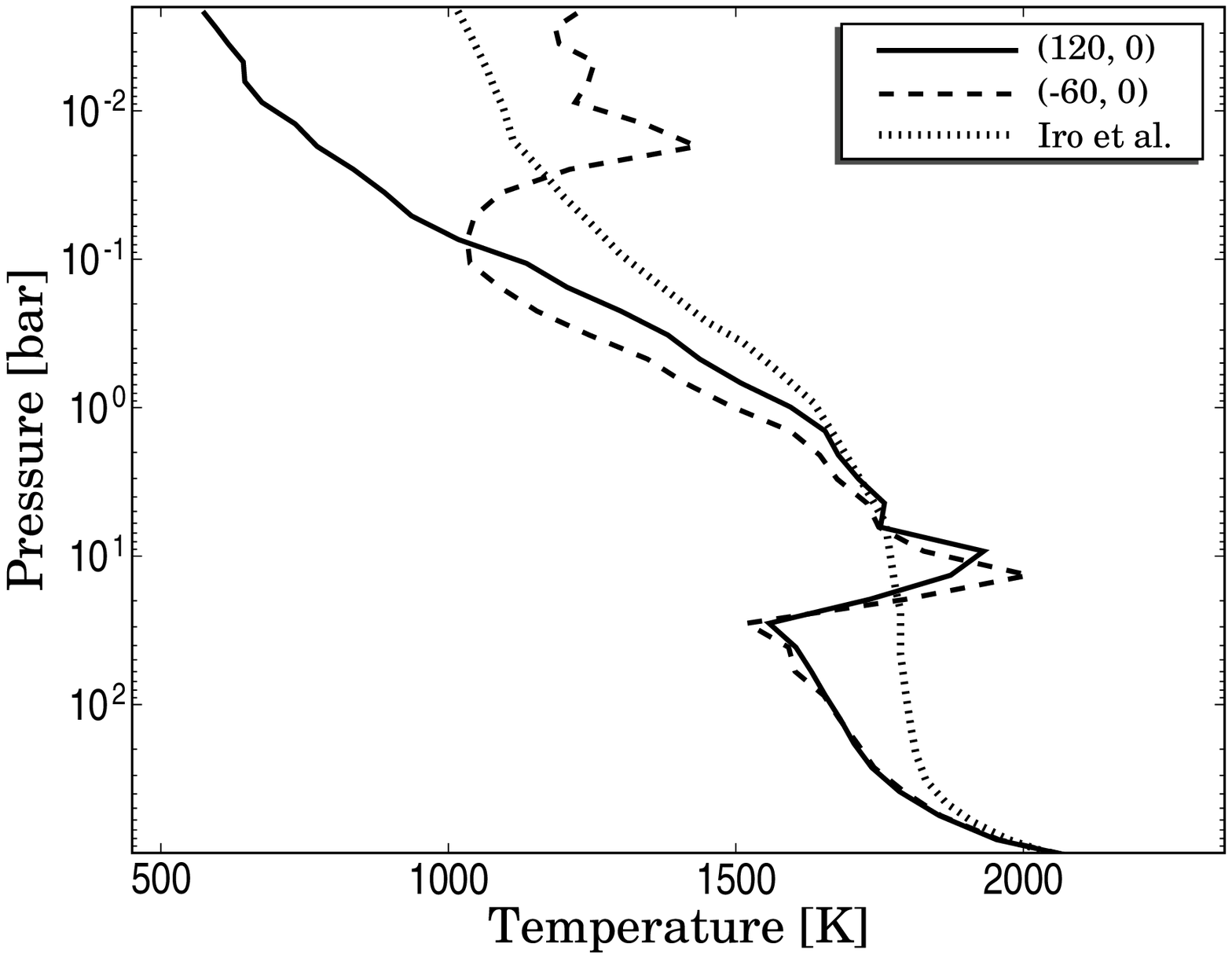}\\[5mm]
\includegraphics[scale=0.45]{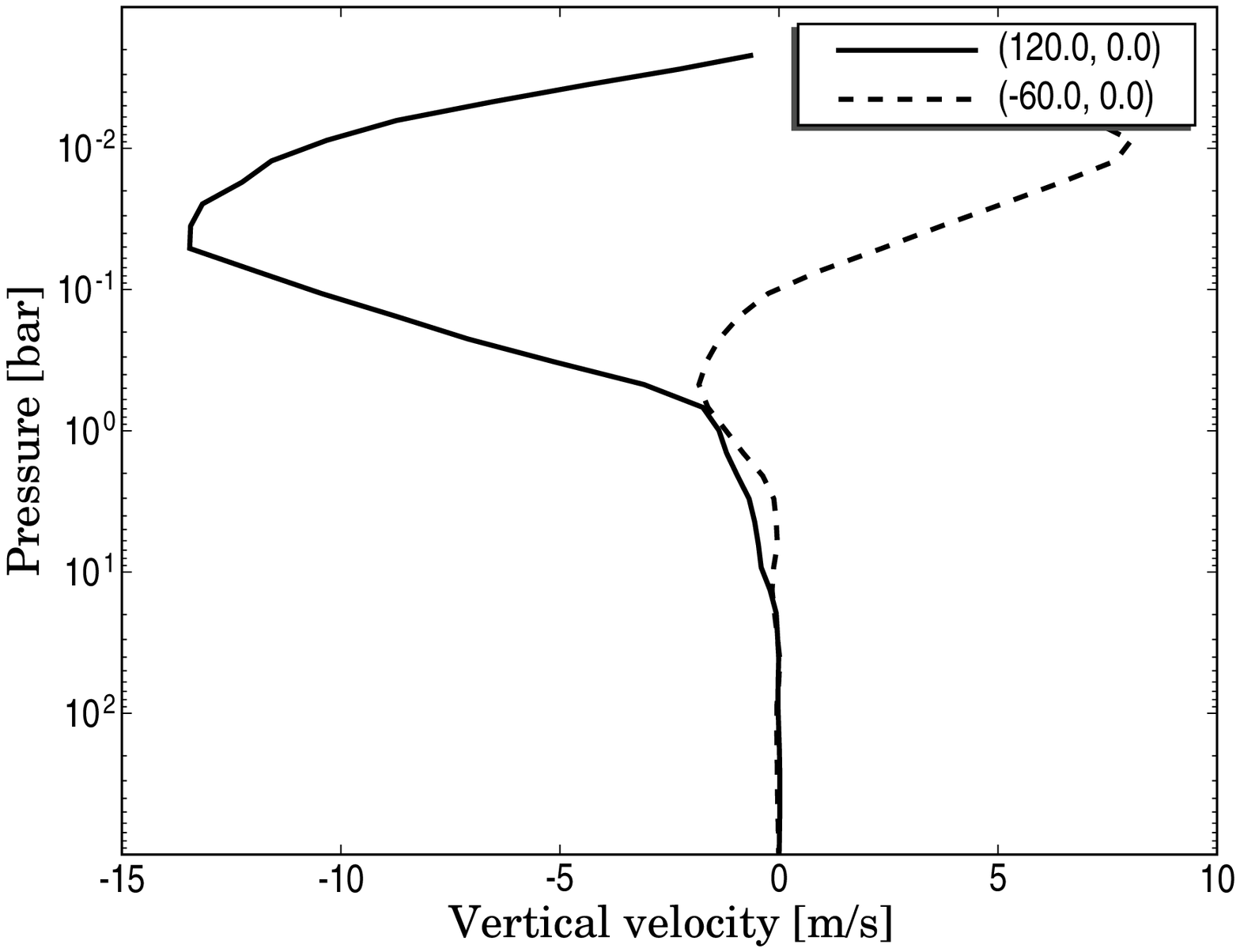}\\
}
\caption{
Demonstrates quenching according to the eddy diffusion model of
\citet{Smith:1998} in the nominal simulation for two vertical columns of
atmosphere.  The top panel shows $\tau_{\rm chem}$ and $\tau_{\rm dyn}$ on two
vertical columns: (1) ($\lambda = 120\degr, \phi = 0\degr$) (solid lines) and
(2) ($\lambda = -60\degr, \phi = 0\degr$) (dashed lines).  The quench level
for the first column is where the thick and thin solid lines cross.  Likewise,
the quench level at ($\lambda = -60\degr, \phi = 0\degr$) is where the thick
and thin dashed lines cross.  The middle and bottom panes show the
temperatures and vertical velocities used to calculate $\tau_{\rm dyn}$ in
each region.
}
\label{figure:quench_ss_as}
\end{figure}

The simplest model of disequilibrium chemistry defines the quench level to be
where the timescale for reactions equals the timescale for dynamical advection
in the system: i.e., quenching occurs at $\rm \tau_{\rm chem} = \tau_{\rm
dyn}$.  The latter quantity, $\tau_{\rm dyn}$, represents the timescale for
dynamical motions to advect material across regions of atmosphere differing
appreciably in temperature and pressure.  Depending on local gradients in
temperature, these timescales can be either large or small.

These considerations led \citet{Smith:1998} to revise the eddy diffusion model
of quenching originally proposed by \citet{Prinn:1977}.  In
\citet{Smith:1998}'s model, the dynamical timescale is determined by eddy
diffusion:
\begin{equation}
\tau_{\rm dyn} = \frac{L}{w},
\label{equation:tau_dyn}
\end{equation}
where $L$ is a length scale and $w$ is the vertical velocity near
the quench level. 

Previous researchers estimated the quench level on Jupiter assuming $L = H$,
where $H$ is the pressure scale height of the atmosphere.  \citet{Smith:1998}
demonstrates that the assumption of $L = H$ is incorrect for most planetary
atmospheres.  However, the eddy diffusion calculation is approximately correct
if $L$ is replaced with an effective length scale $L_{\rm eff}$, ``which
depends on the e-folding length scales of $\tau_{\rm chem}$ and $\tau_{\rm
dyn}$ and the pressure dependence of the equilibrium value of the chemical
abundance.''  He goes on to develop a recipe for determining $L_{\rm eff}$.

We apply the \citet{Smith:1998} recipe, which offers excellent insight into
the chemistry governing our simulations, to the nominal atmosphere model.  We
also assume for the moment that quenching occurs in the vertical direction,
not horizontally.  The crucial quantity needed to estimate eddy diffusion is
the vertical velocity (in height coordinates).  Given $\omega = Dp/Dt$, the
vertical velocity in height coordinates is approximately given by 
\begin{equation}
w = \omega/(-\rho g), 
\end{equation}
where $\rho$ is the density and $g$ is the acceleration of gravity.  It should
be noted that this is not exact but is an excellent approximation for shallow
atmospheres \citep[][p. 77--80]{Holton:1992}.  

\begin{figure}
\includegraphics[scale=0.525]{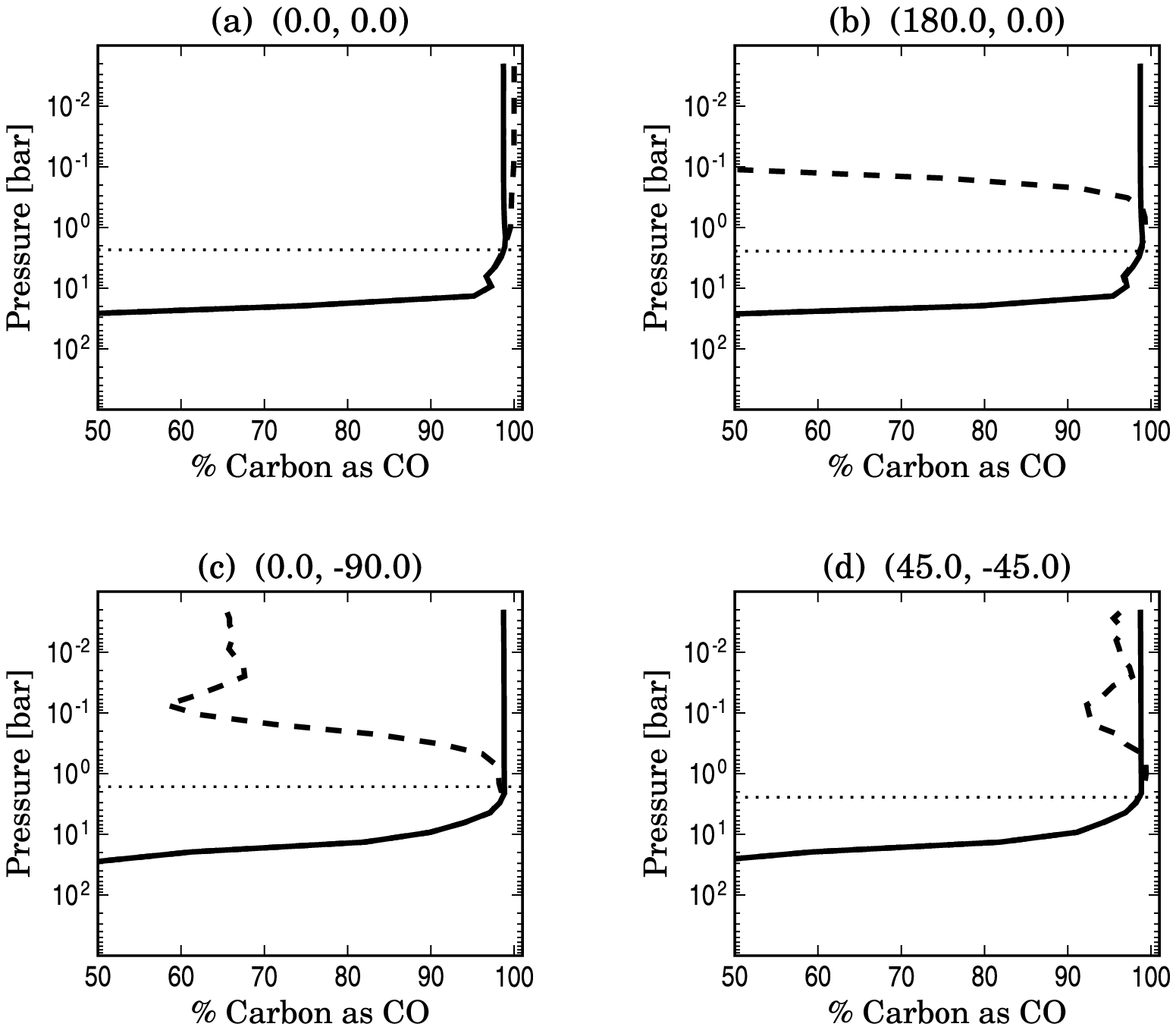}
\caption{
Comparison of actual (solid line) and equilibrium (dashed line) CO abundance
in the nominal simulation.  Plotted is the \% of carbon as CO---as opposed to
$\rm CH_4$.  Titles above the panels give the longitude and latitude for each
column.  The horizontal dotted lines show the quench level for each column as 
predicted with the \citet{Smith:1998} recipe.
}
\label{figure:CO_ss_as}
\end{figure}

To estimate quenching, it is essential to determine the appropriate length
scale $L_{\rm eff}$ to use in computing $\tau_{\rm dyn}$, the timescale for
eddy diffusion.  Following the \citet{Smith:1998} recipe, $L_{\rm eff} \approx
0.5\;H$ in the first region, and $L_{\rm eff} \approx 0.3\;H$ second region.
These are closer to $H$ than are appropriate for Jupiter because the
equilibrium abundance of CO is uniformly high at the equator for $p > 200$
mbar.  This correction nonetheless causes a significant shift in the pressure
at which quenching occurs in each case. 

We show the quench levels (using $\tau_{\rm dyn}$ = $L_{\rm eff}/w$, not
$H/w$) for both profiles in the top panel of Figure \ref{figure:quench_ss_as}.
The quench level near ($\lambda = 120\degr, \phi = 0\degr$) is about 3.0 bars;
near ($\lambda = -60\degr, \phi = 0\degr$), it is at 2.5 bar.  The middle and
bottom panels of Figure \ref{figure:quench_ss_as} show the vertical velocities
and temperatures in these two regions.  Here, we have averaged columns of $w$
over a $45\degr$ angular radius on the sphere, which removes small-scale
variations but preserves the large-scale structure of the vertical wind.  For
$p < 1$ bar, downwelling occurs in the first region, whereas the second
region, which is centered on the opposite hemisphere, has upwelling.

The \citet{Smith:1998} quenching model also predicts a homogeneous
distribution of CO above the quench level, with the value determined by the
equilibrium CO abundance at the quench level.  Since this is high in both
cases, the eddy diffusion model also predicts very little $\rm CH_4$ at the
photosphere of HD 209458b.  To emphasize the agreement between the
\citet{Smith:1998} view of quenching and the actual abundance of CO in our
simulations, we compare the equilibrium view to the CO abundance predicted in
our simulation in Figure \ref{figure:CO_ss_as}.  This has been done on four
columns of atmosphere to show the variation in quenching in different
atmospheric regions.  The \citet{Smith:1998} quench level in each pane is
indicated by a thin dotted line, which agrees well with the convergence of the
equilibrium vs. disequilibrium profiles (thick dashed and solid lines,
respectively).

We conclude this discussion by reiterating why horizontal quenching near the
day-night terminator is not of great importance in the simulation.  Using
reasoning analogous to our discussion above, the horizontal mixing timescale
should be $\tau_{\rm dyn,\,h} = L_{\rm h}/u$, where $u$ is the supersonic
zonal wind and $L_{\rm h}$ is a length scale appropriate for horizontal
quenching.  Lacking better information, $L_{\rm h} = 2\pi R_{\rm p}$ (just the
circumference of the planet).  Note that the true value of the horizontal
mixing time may be smaller than $2\pi R_{\rm p}/u$ in regions where parcels
travel through large temperature gradients.  But given the extremely fast wind
speeds ($u\sim\!3000\;\rm m\;s^{-1}$) predicted in the upper layers,
$\tau_{\rm dyn,\,h}$ can never be larger than this value.  Assuming for the
moment that this is correct, we obtain $\rm \tau_{dyn,\,h} \sim\! 2 \times
10^5\;s$.  But Figure \ref{figure:quench_ss_as} clearly shows that $\tau_{\rm
chem} \gg \rm 10^5\;s$ for $p < 3$ bar, where vertical quenching is achieved.
Hence, the quench level in the vertical has already been reached at levels
where $\tau_{\rm chem} < \tau_{\rm dyn,\,h}$.

\section{Discussion}
\label{section:Discussion}

\subsection{Role of Photochemistry}
\label{subsection:Photochemistry}

\citet{Liang:2004} show that the distribution of photochemical products,
including aerosols and radicals, is insignificant at the pressure levels
considered here compared to the primary constituents ($\rm H_2$, He, $\rm
H_2O$, CO, and possibly $\rm CH_4$).  But the effects of these species on the
$\rm CO/CH_4$ reactions has not been explored here.  Conceivably, OH and H
coming from the photolysis of $\rm H_2O$ \citep{Liang:2003} could influence
the reaction rates of the \citet{Yung:1988} sequence.  

We have not explored the consequences of photochemically enhanced OH and H
populations on the $\rm CO/CH_4$ interconversion chemistry here.  This effect
is most likely to be noticeable in the highest layers of the model ($p$ < 10
mbar), above which the stellar ultra-violet (UV) light is absorbed.  In order
to have an important effect on our predictions for the CO distribution,
however, the chemical interconversion timescale resulting from interaction
with photochemical radicals would have to be reduced to $\tau_{\rm chem} =
10^6\;\rm s$ (or less).  Such timescales would be very short compared with the
values listed in Table \ref{table:tau_chem} at these temperatures and
pressures.  That is, unless the photochemical production of radicals dominates
the chemistry, the disequilibrium distributions will be substantially similar
to those presented here.

\subsection{Treatment of Radiation}
\label{subsection:Radiation}

Asynchronous rotation on hot Jupiters is possible, either due to
eccentricity/obliquity pumping by other planets in the system or thermal tides
\citep{Showman:2002}.  In the case of HD 209458b, we expect synchronous
rotation to hold more closely than planets in eccentric orbits.  An
asynchronous component to the rotation, due to thermal tides and transfer of
angular momentum from the atmosphere to the interior, could conceivably
influence the atmospheric dynamics solution (Figure
\ref{figure:T_winds_nominal}).  For example, our time-steady flow scenario
might be superposed in some regions by a slow periodicity resulting from a
radiative heating profile that slowly advances along the meridians.  But as we
discuss in \S \ref{section:Analysis_Interpretation}, quenching is in the
vertical.  Adding asynchronous rotation to the model might change the details
of the wind and temperature patterns but not the mean vertical velocity.
Hence, the $\rm CO/CH_4$ distributions should not be sensitive to this
correction.

Newtonian heating ignores potentially subtle feedbacks in the radiation
physics.  It has been used successfully, however, to model essential features
of Earth's general circulation \citep[][p. 409--411]{Holton:1992}.  We note
that the approximation is likely to be less valid deep in the atmosphere of
EGPs than it is for Earth.  This is due to long radiative time constants and
huge departures from radiative equilibrium.  Hence, the use of Newtonian
heating is probably the major source of error in our atmosphere model.  We use
it here partly because the approximation is reasonably valid for $p < 2$ bar,
where $\tau_{\rm rad}$ is less than several days.  The scheme is also
relatively easy to implement and performs well computationally, which is very
desirable, as the AGDC2 itself demands considerable computational resources.
We are able, therefore, to test the sensitivity of the results to a broad set
of input parameters.

We note also that the abundances of $\rm H_2O$, CO, and $\rm CH_4$---used by
\citet{Iro:2005} to derive the contributions of these molecules to the total
opacity---are calculated based on chemical equilibrium models.  We demonstrate
here, however, that chemical equilibrium does \emph{not} hold for $p \ll 10$
bars.  This internal inconsistency is inherent in our use of Newtonian cooling
and is unfortunately unavoidable.  Radiative-equilibrium models of HD
209458b's atmosphere to date have all assumed chemical equilibrium.  Even in
the future, this coupling between dynamics and radiation will be difficult to
disentangle because, as has been previously discussed, disequilibrium
abundances depend on the overall meteorology.

\citet{Iro:2005} also do not consider clouds.  Where present, high silicate
clouds could dominate the opacity and greatly perturb the atmospheric profile.
Cloud deck optical thickness in general varies with wavelength, depending on
the vertical extent of the cloud and the condensate particle sizes
\citep{Ackerman:2001, Cooper:2003}.  It is unclear what properties a silicate
cloud high in the atmosphere would have.  This is a complicated subject
involving intricate feedbacks between radiation, cloud microphysics, and
dynamics.  We neglect the possibly crucial effects of clouds here.  Including
condensation processes will be a major future advance in the field of EGP
atmosphere modeling.

\subsection{Initial Conditions}
\label{subsection:Initial_Conditions}

Given the large $\tau_{\rm chem}$ values at low pressures, it is reasonable to
wonder whether our solutions for the CO distribution depend on the initial
conditions.  For the simulations shown, we have assumed that CO is initially
in chemical equilibrium everywhere (equation \ref{equation:X_CO}).  As
discussed in \citet{Cooper:2005}, temperatures and winds are not strongly
sensitive to the initial conditions.  The primary concern here is whether
vertical transport can mix CO upward rapidly enough to bring the CO mole
fraction at the photosphere up to its value at the quench level.  

Clearly, chemical processes occurring on timescales longer than $\sim\!1$ year
cannot have run to completion by 1000 days of integration time.  But we can
argue on simple grounds that 1000 days is still long enough to determine
$X_{\rm CO}$, even where $\tau_{\rm chem} \gg 1$ year.  Assuming the vertical
wind at 1 bar is $\sim\!10\rm\;m\;s^{-1}$, as shown by the model, and the
vertical depth from the photosphere (100 mbar) to the quench level (3 bar) is
$\sim\!2$\% of the planetary radius (several scale heights), we would expect
upward mixing to occur on timescales of $\sim\!1$ week.  Thus, after
$\sim\!1000$ days of integration time, the CO concentration should approach
the value at the quench level.

We have run two additional simulations with different initial conditions.  The
first simulation assumes no initial CO in the atmosphere; i.e., 100\% of the
carbon is $\rm CH_4$.  The second simulation assumes no initial $\rm CH_4$;
i.e., 100\% of the carbon is CO.  Both use the input parameters of the nominal
model in all other respects.  Therefore, temperatures and winds are the same
as in the nominal case.  After even several hundred days of simulation time,
the solutions for CO in these two cases are virtually indistinguishable from
the nominal simulation (see Figure \ref{figure:tracers_nominal}).  These tests
demonstrate that memory of the initial condition for $X_{\rm CO}$ has been
erased by $t_{\rm sim} = 1000$ days.

\subsection{Metallicity and C/O}
\label{subsection:Metallicity}

Our nominal simulation assumes solar abundances based on the
\citet{Lodders:1998} compilation.  We note that the precise values of the
solar abundances, especially C, N, and O, have undergone minor modifications
in the past seven years, some of which are still under discussion within the
scientific community \citep{Grevesse:1998, Mahaffy:2000, Prieto:2001,
Prieto:2002, Lodders:2003, Asplund:2006}.  In particular, the CNO abundances,
along with those of the noble gases, have been revised down from the
geochemically based abundances of \citet{Anders:1989, Lodders:1998}.  The C/O
ratio is now thought to be exactly 0.5; following \citet{Lodders:1998}, we use
C/O = 0.57 here.  Incorporating these updates to the solar abundances will
modify the results presented here somewhat, though the correction is likely
to be fairly minor.

From stellar spectroscopy, the endowment of heavy elements in the star HD
209458 is well-known to be about solar: $\rm [Fe/H] = 0.014 \pm 0.03$
\citep{Fischer:2005}.  That says nothing about the metallicity of the planet
orbiting it, which has not yet been measured.  Jupiter and Saturn boast
pronounced enrichments in heavy elements relative to the Sun, suggesting EGPs
may also be metal-rich compared to their host stars.  We consider here the
effects of increasing the relative proportions of carbon and oxygen in HD
209458b's atmosphere.  

Suppose the constants $c_1$ and $c_2$ in equations \ref{equation:c1} and
\ref{equation:c2} are both increased by a factor of 10.  This is a much
greater metal abundance than is seen on Jupiter, which has enrichment of
carbon and nitrogen by a factor of $\sim\!3$ relative to the Sun \citep[from
Galileo probe measurements; the probe was not successful in measuring the
abundance of oxygen---see][]{Wong:2004}.  As carbon and oxygen in this case
are still minor constituents, this change will not appreciably affect the mean
molecular weight or heat capacity of the atmosphere.  According to our
chemical equilibrium solution, equation \ref{equation:X_CO}, increased [Fe/H]
favors the formation of CO.  That is, for a given temperature and pressure,
the ratio $X_{\rm CO}/X_{\rm CH_4}$ is \emph{higher} for [Fe/H] = +1.0 than it
is at [Fe/H] = 0.0, in agreement with \citet{Lodders:2002}.  This implies
that, relative to the solar metallicity case, the quench level should have a
greater abundance of CO in the [Fe/H] = +1.0 atmosphere by \emph{more} than a
factor of 10.  Hence, we would also expect the photosphere to show highly
abundant CO.

We have run a high-metallicity simulation, in which we set [Fe/H] = 1.0 in the
atmosphere.  The other parameters of this model are identical to the nominal
simulation, so the dynamics are the same.  We indeed see in this simulation
that 100\% of the carbon atoms are in CO, with no $\rm CH_4$ at all (as
opposed to 1--2\% $\rm CH_4$ predicted in the nominal case).  

We stress that the above discussion assumes atmospheric temperatures do not
change drastically as a result of increasing the metallicity.  This assumption
may not be correct for all EGPs.  We note, however, that \citet{Fortney:2005a}
consider metallicity adjustments to their radiative-equilibrium models of HD
209458b and TrES-1.  They find that the effective temperature of TrES-1
increases by only $\sim\!14\;\rm K$ for a metallicity of $5\times$ solar.  The
direction of the effect---warming, not cooling---also favors the appearance of
CO.

Similarly, increasing the C/O ratio in the atmosphere will not inhibit CO's
formation at the temperature of HD 209458b's atmosphere.  In the Sun, $\rm C/O
\approx 0.5$.  If $\rm C/O \ga 1$, CO can potentially take up 100\% of the
available oxygen, in which case $\rm H_2O$ will be depleted.  Atmospheres
showing the spectral signatures of CO without prominent $\rm H_2O$ features
very likely have a high C/O ratio relative to solar.  The effects of
super-solar abundances and enhanced C/O are discussed in more detail elsewhere
\citep{Fortney:2005a, Seager:2005}.

\section{Conclusions / Implications}
\label{section:Implications}

We have run 3D numerical simulations of HD 2094548b's atmosphere to explore
the chemistry of carbon.  We advect CO through the atmosphere as a passive
tracer, relaxing its concentration toward the local chemical equilibrium
value.  The relaxation timescale depends on the reaction pathways that convert
CO to $\rm CH_4$.  We follow recent previous work on disequilibrium carbon
chemistry \citep{Griffith:1999, Bezard:2002} and assume that the
\citet{Yung:1988} sequence is the most efficient mechanism for the chemical
reduction of CO.  

We calculate the relative abundances of CO, $\rm CH_4$, and $\rm H_2O$, which
are the main species of carbon and oxygen in the atmosphere.  Our nominal
simulation models HD 209458b specifically.  For comparison, we have also run
simulations of ``cold'' and ``hot'' atmospheres.  The cold simulation is
representative of a planet that is $\sim\!300$ K cooler than HD 209458b; the
hot simulation models a planet $\sim\!300$ K warmer.  

The atmospheric dynamics by 1000 Earth days of simulation time has reached the
steady state described in \citet{Cooper:2005}.  Eastward winds in the nominal
model reach $\rm \sim\!4000\;m\;s^{-1}$ at the photosphere, with corresponding
temperature contrasts of $\rm \sim\!500\;K$.  Atmospheric dynamics partially
redistribute the thermal energy to the planet's nightside, but longitudinal
contrasts in temperature are apparent down to $\sim\!2$ bars.

Chemical equilibrium has been the basis of all prior discussions about CO and
$\rm CH_4$ on HD 209458b.  We show here, though, that the equilibrium picture
of $\rm CO/CH_4$ chemistry is a vast misconception.  In regions where chemical
equilibrium does not hold, the true CO abundance depends not only on local
quantities but also on the overall meteorology.  

We show that the timescale for $\rm CO/CH_4$ interconversion depends
sensitively on both temperature and pressure (Table \ref{table:tau_chem}).  At
low temperatures and pressures, reactions are slow; at high temperatures and
pressures, reactions are fast.  As a result of long reaction times, CO is not
expected to be in chemical equilibrium in the 1--1000 mbar region (the
``visible'' layers).  Chemical equilibrium in the model atmosphere is attained
deeper in the atmosphere ($p \ga 5$ bars), where the interconversion
timescale is short relative to dynamical mixing times.  

The atmosphere is well-mixed by vertical motions, which leads to vertical (as
opposed to horizontal) quenching.  Vertical quenching results in photospheric
CO concentrations exceeding chemical equilibrium values by many orders of
magnitude.  The simulations show that CO, $\rm CH_4$, and $\rm H_2O$ will be
homogeneously distributed throughout the photosphere of HD 209458b and similar
planets.  At $p \la 2$ bars, almost all of the carbon is in the form of CO,
even in cold regions such as the nightside.  Consequently, we expect water
vapor to be less plentiful than previously thought.  These conclusions are
consistent with the simpler view of quenching presented by \citet{Smith:1998}.  

The metallicity of HD 209458b is not known.  Our nominal simulation assumes
solar abundances.  We have also run a variation on the nominal model that
assumes ten times enhancement of the heavy elements: [Fe/H] = +1.0.
Disequilibrium effects operate similarly in this model, and the ratio of CO to
total carbon in the atmosphere remains close to 100\%.  Indeed, higher
metallicity favors the formation of CO over $\rm CH_4$.  Hence, if close-in
EGPs have characteristically super-solar abundances, CO could be prominent at
the photospheres of planets even cooler than $\rm T_{eff} = 1000\;K$.

Likewise, increasing the C/O ratio would not significantly inhibit CO's
formation at depth (and subsequent transport up to the photosphere).  If C/O
exceeds the solar value of 0.5 in the Sun, CO on HD 209458b should be highly
abundant.  The tell-tale sign of this scenario at HD 209458b's effective
temperature is $\rm H_2O$ depletion.

Our results have several interesting consequences for near-future observations
of close-in EGPs, although a thorough analysis of close-in EGP spectra is
beyond the scope of this work.  We conclude simply with a brief description of
the implications of these findings.

The hot model shows that planets significantly warmer than HD 209458b should
have \emph{no} $\rm CH_4$ in their photospheres, unless the C/O ratio is
greater than 1.  But in planets significantly cooler than HD 209458b,
disequilibrium effects should raise the concentration of CO to detectable
levels.  The specific value of $X_{\rm CO}$ cannot be predicted with definite
precision based on these simulations; there are too many uncertainties in the
model's inputs.  However, examination of the values of $\tau_{\rm chem}$
(Table \ref{table:tau_chem}) suggests that 1--10 bar may be the approximate
quench level for \emph{all} close-in EGPs within $\rm 750\;K$ of HD 209458b's
effective temperature ($\rm T_{eff}\sim\!1350$ K for this planet).  The
relative concentrations of CO and $\rm CH_4$ at the photosphere are therefore
diagnostic of local conditions in the 1--10 bar region of the atmosphere,
which is deeper than observations can directly probe.  Neither species has yet
been detected in this planet's atmosphere \citep{Richardson:2003a,
Richardson:2003b, Deming:2005b}.  Our results here suggest that $\rm CH_4$
depletion (not depletion of CO) is the more likely situation; even in
chemical equilibrium, $\rm CH_4$ is likely to be depleted at the limb.

Measurements of carbon-bearing species can thus serve as a useful calibration
tool of radiation and dynamics models of these planets.  It should be noted
that metallicity also plays a role in determining the abundance of CO at the
quench level.  This may be a difficult effect to deconvolve from temperature
and pressure, unless absolute concentrations can be measured.  For near-future
explorations, what is crucial here is that planets with little or no $\rm
CH_4$ must be very much warmer than TrES-1 or else cooler, but very metal-rich
(as high metallicity favors the formation of CO over $\rm CH_4$).

It is known that HD 209458b does contain carbon and oxygen
\citep{Vidal-Madjar:2004}.  As \citet{Deming:2005b} discuss, the null
detection of CO using NIRSPEC on Keck II cannot be easily attributed to CO
depletion in HD 209458b's atmosphere.  This lends credence to the hypothesis
of a high cloud or haze obscuring the spectral features of CO near 2
$\micron$.  As discussed by \citet{Fortney:2005b}, however, the slant geometry
implies that even a layer of modest optical thickness could hide the signature
of CO.  It may be possible to distinguish between these and other scenarios.  

Our results suggest that TrES-1, which orbits a K-type star, should also show
strong CO absorption features in its spectrum.  A high silicate condensation
cloud obscuring the photosphere is not expected on TrES-1 due to lower
temperatures overall.  The condensation of silicates should occur at much
greater depth in TrES-1's atmosphere.  It would be very illuminating to
perform a transmission spectroscopy search for CO on TrES-1.  Discovery of CO
on TrES-1 would make a strong case for high silicate clouds in the atmosphere
of HD 209458b.  If CO is also not detectable on TrES-1 by transmission
spectroscopy, on the other hand, the hypothesis of stratospheric haze (that
hides CO on both planets) becomes more appealing.

The ubiquitousness of CO throughout the photospheres of close-in EGPs should
also have profound implications for the spectra of these objects.  Our results
for the cold model suggest that CO may even be abundant on cooler EGPs than
TrES-1.  If so, CO's spectral signature should be detectable.  The absorption
features of $\rm H_2O$ may also be less pronounced, especially on hot planets,
in which over half of the available oxygen atoms are bound up in CO.  In
addition to the 2 $\micron$ band pass explored by \citet{Deming:2005b}, we
suggest searches for CO in the rovibrational fundamental near 4.5 $\rm
micron$.  If present, the opacity of CO overwhelms that of $\rm H_2O$ vapor
at this wavelength, even more so than the wavelength region explored in
\citet{Deming:2005b}.  As an IR window also exists at 4.5 $\micron$, this
observation may also be possible using ground-based platforms.

It is noteworthy also that these simulations predict a homogeneous
distribution of CO, $\rm CH_4$, and $\rm H_2O$ in HD 209458b's atmosphere,
although the temperature is almost certainly not isothermal at the
photosphere.  \citet{Cooper:2005} show that temperature variations over the
photosphere can lead to brightness variations in the planet with orbital
phase.  It must be emphasized that the synthetic light curve of
\citet{Cooper:2005} pertained to the \emph{bolometric} luminosity of the
planet (i.e., the emission integrated over all wavelengths); nothing is
directly implied about the planet's brightness at a particular wavelength.
The flux contrast of the planet's photosphere over a full orbital period would
likely be suppressed somewhat by a homogeneous distribution of CO and $\rm
CH_4$, as the distribution of these species determines the level of the
photosphere [J. Fortney, private communication].  Wavelength-dependent light
curves will require generating synthetic spectra using a radiative transfer
model, a task we leave for future work.  But we hope this discussion will
inspire near-future searches for CO and $\rm CH_4$ on these planets, some of
which should be possible even with ground-based telescopes.


\acknowledgements

Special thanks to J. I. Lunine, J. J. Fortney, J. W. Barnes, C. A. Griffith,
B.  B\'ezard, J. I. Moses, M. A. Pasek, B. K. Jackson, and R. V. Yelle for
valuable ideas and insights during the course of this project and guidance in
developing our treatment of the carbon chemistry.  Thanks you to the referee
for helpful comments in improving the manuscript.
Figures created using the free Python SciPy, Matplotlib, and PLplot libraries.
Thanks also to Josh Bernstein, John Pursch, Tony Farro, and Joe Gotobed for
their hard work in configuring and maintaining the LPL Beowulf-style cluster,
which these simulations ran on for weeks.  This research was supported by NSF
grant AST-0307664 and NASA GSRP NGT5-50462.

\newpage
\bibliographystyle{apj}


\begin{deluxetable}{llllllllll}
\tabletypesize{\scriptsize}
\tablecaption{CO in Chemical Equilibrium \label{table:CO_fraction}}
\tablehead{ \colhead{T [K]} & \colhead{10 mbar} & \colhead{100 mbar} & \colhead{500 mbar} 
& \colhead{1 bar} & \colhead{5 bar} & \colhead{10 bar} 
& \colhead{100 bar} & \colhead{500 bar} & \colhead{1 kbar} }
\tablewidth{0pt}
%
\startdata
500 & $1.7\times 10^{-9}$ & 0.0 & 0.0 & 0.0 & 0.0 & 0.0 & 0.0 & 0.0 & 0.0	    \\
750 & $5.5\times 10^{-2}$ & $6.0\times 10^{-4}$ & $2.4\times 10^{-5}$ & $6.0\times 10^{-6}$ 
    & $2.4\times 10^{-7}$ & $6.0\times 10^{-8}$ & $6.3\times 10^{-10}$ & 0.0 & 0.0	    \\
1000 & 1.0 & 0.73 & 0.15 & $4.4\times 10^{-2}$ & $1.9\times 10^{-3}$ & $4.6\times 10^{-4}$ 
    & $4.7\times 10^{-6}$ & $1.9\times 10^{-7}$ & $4.7\times 10^{-8}$	    \\
1250 & 1.0 & 1.0 & 0.95 & 0.85 & 0.27 & $9.4\times 10^{-2}$ & $1.1\times 10^{-3}$ 
    & $4.4\times 10^{-5}$ & $1.1\times 10^{-5}$     \\
1500 & 1.0 & 1.0 & 1.0 & 0.99 & 0.89 & 0.71 & $4.0\times 10^{-2}$ 
    & $1.7\times 10^{-3}$ & $4.2\times 10^{-4}$	    \\
1750 & 1.0 & 1.0 & 1.0 & 1.0 & 0.99 & 0.96 & 0.32 & 
    $2.2\times 10^{-2}$ & $5.7\times 10^{-3}$	    \\
2000 & 1.0 & 1.0 & 1.0 & 1.0 & 1.0 & 0.99 & 0.70 & 0.13 & $3.7\times 10^{-2}$	    \\
2500 & 1.0 & 1.0 & 1.0 & 1.0 & 1.0 & 1.0 & 0.96 & 0.61 & 0.32
\enddata
%
\tablecomments{Fraction of carbon in the atmosphere present as CO (i.e., the
ratio $X_{\rm CO}\,/\,c_1$), assuming chemical equilibrium (see eq.
\ref{equation:c1} -- \ref{equation:X_CO}).  We obtain these values by solving
equation \ref{equation:X_CO}, which calculates the chemical equilibrium of
CO in a gas of solar composition.  Our treatment follows \citet{Lodders:2002}.
Data for the standard Gibbs free energies of formation for CO, $\rm CH_4$, and
$\rm H_2O$ are taken from the NIST-JANAF tabulations \citep{Chase:1998}.
}
\end{deluxetable}

\begin{deluxetable}{llllllllll}
\tabletypesize{\scriptsize}
\tablecaption{ CO/$\rm CH_4$ interconversion timescale \label{table:tau_chem} }
%
\tablehead{ \colhead{T [K]} & \colhead{10 mbar} & \colhead{100 mbar} &
\colhead{500 mbar} & \colhead{1 bar} & \colhead{5 bar} & \colhead{10 bar} &
\colhead{100 bar} & \colhead{500 bar} & \colhead{1 kbar} }
%
\tablewidth{0pt}
%
\startdata
500 & $1.0\times 10^{24}$ & $4.3\times 10^{21}$ & $1.7\times 10^{20}$ & $4.9\times 10^{19}$ & $3.7\times 10^{18}$ 
    & $1.3\times 10^{18}$ & $4.0\times 10^{16}$ & $3.5\times 10^{15}$ & $1.3\times 10^{15}$ \\
750 & $9.0\times 10^{17}$ & $2.8\times 10^{15}$ & $5.2\times 10^{13}$ & $9.3\times 10^{12}$ & $1.9\times 10^{11}$
    & $3.9\times 10^{10}$ & $4.2\times 10^{8}$ & $3.1\times 10^{7}$ & $1.1\times 10^{7}$ \\
1000 & $1.1\times 10^{15}$ & $3.6\times 10^{12}$ & $6.4\times 10^{10}$ & $1.1\times 10^{10}$ & $2.1\times 10^{8}$
    & $3.7\times 10^{7}$ & $1.4\times 10^{5}$ & $4.8\times 10^{3}$ & $1.3\times 10^{3}$  \\
1250 & $2.3\times 10^{13}$ & $7.4\times 10^{10}$ & $1.3\times 10^{9}$ & $2.3\times 10^{8}$ & $4.2\times 10^{6}$
    & $7.4\times 10^{5}$ & $2.4\times 10^{3}$ & 52 & 11 \\
1500 & $1.9\times 10^{12}$ & $6.0\times 10^{9}$ & $1.1\times 10^{8}$ & $1.9\times 10^{7}$ & $3.4\times 10^{5}$
    & $6.0\times 10^{4}$ & $1.9\times 10^{2}$ & 3.6 & 0.68 \\
1750 & $3.3\times 10^{11}$ & $1.0\times 10^{9}$ & $1.9\times 10^{7}$ & $3.3\times 10^{6}$ & $5.9\times 10^{4}$
    & $1.0\times 10^{4}$ & 33 & $0.60$ & $0.11$	\\
2000 & $9.1\times 10^{10}$ & $2.9\times 10^{8}$ & $5.1\times 10^6$ & $9.1\times 10^{5}$ & $1.6\times 10^{4}$
    & $2.9\times 10^{3}$ & 9.1 & 0.16 & $2.9\times 10^{-2}$ \\
2500 & $1.6\times 10^{10}$ & $5.0\times 10^{7}$ & $8.9\times 10^5$ & $1.6\times 10^{5}$ & $2.8\times 10^{3}$
    & $5.0\times 10^{2}$ & 1.6 & $2.8\times 10^{-2}$ & $5.0\times 10^{-3}$
\enddata
%
\tablecomments{ Timescale in seconds for CO/$\rm CH_4$ interconversion in the
\citet{Yung:1988} catalyzed reaction sequence at the same temperatures and
pressures used in Table \ref{table:CO_fraction}.  These values have been
computed using equations \ref{equation:tau_chem2} and
\ref{equation:K_CH3O}, following equations (7) and (8) of
\citet{Bezard:2002}.  The expression for $\rm \tau_{chem}$ derives from
consideration of the rate of the reverse of reaction \ref{reaction:Yung1},
which has been measured in the laboratory \citep{Page:1989}.
}
\end{deluxetable}

\end{document}